\documentclass[twocolumn,showpacs,preprintnumbers,amsmath,amssymb]{revtex4}
\usepackage{graphicx}
\usepackage{dcolumn}
\usepackage{bm}
\usepackage{here}

\begin{document}
\draft

\title{Exactly solvable scale-free network model}

\author{Kazumoto Iguchi\cite{byline} and Hiroaki Yamada\cite{byline2}}

\address{{\it 70-3 Shinhari, Hari, Anan, Tokushima 774-0003, Japan\/}\cite{byline}}
\address{{\it 5-7-14 Aoyama, Niigata 950-2002, Japan\cite{byline2}\/}}

\date{\today}

\begin{abstract}
We study a deterministic scale-free network recently proposed by Barab\'{a}si, Ravasz and Vicsek.
We find that there are two types of nodes: the hub and rim nodes,
which form a bipartite structure of the network.
We first derive the exact numbers $P(k)$ of nodes with degree $k$ for the hub and rim nodes
in each generation of the network, respectively.
Using this, we obtain the exact exponents of the distribution function $P(k)$ of nodes with $k$ degree
in the asymptotic limit of $k \rightarrow \infty$.
We show that the degree distribution for the hub nodes exhibits the scale-free nature,
$P(k) \propto k^{-\gamma}$ 
with 
$\gamma = \ln3/\ln2 = 1.584962$,
while the degree distribution for the rim nodes is given by
$P(k) \propto e^{-\gamma'k}$
with
$\gamma' = \ln(3/2) = 0.405465$.
Second, we analytically calculate the second order average degree of nodes, $\tilde{d}$.
Third, we numerically as well as analytically calculate 
the spectra of the adjacency matrix $A$ 
for representing topology of the network.
We also analytically obtain the exact number 
of degeneracy at each eigenvalue in the network.
The density of states (i.e., the distribution function of eigenvalues) 
exhibits the fractal nature with respect to the degeneracy.
Fourth, we study the mathematical structure of the determinant 
of the eigenequation for the adjacency matrix.
Fifth, we study hidden symmetry, zero modes and its index theorem 
in the deterministic scale-free network.
Finally, we study the nature of the maximum eigenvalue 
in the spectrum of the deterministic scale-free network.
We will prove several theorems for it, using some mathematical theorems.
Thus, we show that most of all important quantities in the network theory
can be analytically obtained in the deterministic scale-free network 
model of Barab\'{a}si, Ravasz and Vicsek.
Therefore, we may call this network model the exactly solvable scale-free network.
\end{abstract}

\maketitle


\section{Introduction}
There has been a notable progress in the study of the so-called 
scale-free network (SFN)\cite{FFF,Barabasi,AB,KRL,KRR,DMS,DM,KK,BE,BCK}
for recent years.
In the network theory, 
the random network model was
first invented by  Erd\"{o}s and R\'{e}nyi\cite{ER}.
Recently it was generalized to the small world network models
\cite{Klein,WS,NMWS,CNSW,NSW,BA,ASBS,LCS,Strog}.
Furthermore, about five years ago, the SFN was discovered by studying the network
geometry of the internet\cite{FFF,Barabasi,AB,LG,HPPL,AJB1,AJB2}.
Faloutsos brothers\cite{FFF} and 
Albert, Jeong and Barab\'{a}si\cite{Barabasi,AB,LG,HPPL,AJB1,AJB2}
first showed the scale-free nature
of the internet geometry and
opened up an area for studying
very complex and growing network systems such as 
internet, biological evolution, metabolic reaction, epidemic disease,
human sexual relationship, and economy.
These are nicely summarized in the reviews by Barab\'{a}si\cite{Barabasi}.

As was studied in the literature\cite{Barabasi}, 
the nature of these SFNs is characterized by
the power-law behavior of the distribution function.
Here the number of nodes with order $k$ can be fit by 
$$P(k) \propto k^{-\gamma}, \eqno{(1)}$$
where $\gamma \approx 1-4$.
In order to show the power-law distribution of the SFN,
Albert and Barab\'{a}si first proposed a very simple model called
the Albert-Barab\'{a}si (AB)'s SFN model\cite{Barabasi,AB,LG,HPPL,AJB1,AJB2}.

This system is constructed by the following process:
Initially we put $m_{0}$ nodes as seeds for the system.
Every time a new node is added $m$ new links are
distributed from the node to the existed nodes in the system
with a preferential attachment probability
$$\Pi_{i}(k_{i})=\frac{k_{i}}{\sum_{i=1}^{N-1}k_{i}}, \eqno{(2)}$$
where $k_{i}$ is the number of links at the $i$-th node. 
The development of this model is described 
by a continuum model
$$\frac{dk_{i}}{d\tau} = m\Pi_{i}(k_{i}) = \frac{m k_{i}}{2\tau}.  \eqno{(3)}$$
Then at time $\tau$ the system consists of $N(\tau)$ nodes and the $L(\tau)$ links 
with 
$$L(\tau)= \frac{1}{2}\sum_{i=1}^{N(\tau)}k_{i}.  \eqno{(4)}$$
As studied by Barab\'{a}si et. al.\cite{Barabasi}
this model exhibits an exact exponent of 
$\gamma = 3$ 
for the power-law.
Thus, it has been concluded that
the essential points of why a network grows to a SFN
are attributed to the growth of the system and the  preferential attachment 
of new nodes to old nodes existed already in the network.

From the above context,
time evolution to construct an SFN has been intensively studied 
in the $AB$-model as well as other models.
And many works have appeared, regarding nodes and links as metaphysical objects 
such as agents and relationships in an area of science\cite{Barabasi}.
However, most approaches were based on the numerical approach.
And the spectra of the adjacency matrix $A$ for the SFN
have been studied numerically\cite{FFF,Barabasi,FDBV,GKK,DGMS}.
Therefore, apart from the purposes for the numerical analysis, 
the continuous-time SFN models such as the $AB$-model are not good enough to see 
what is going on in the network geometry in the microscopic level.

Instead of such a continuous-time SFN model,
a new type of the SFN models, 
sometimes called deterministic or hierarchical SFN models,
has been proposed by
Barab\'{a}si, Ravasz, and Vicsek\cite{BRV} and Ravasz and Barab\'{a}si\cite{RB}
(We would like to call it the DSFN in this paper).
In the former, the study showed a power-law behavior of the network analytically,
while in the latter, the study showed it numerically.
However, much is still lack and left unanswered.

On the other hand, there is a very important problem on the maximum
eigenvalue, $\lambda_{max}$, of the adjacency matrix, $A$.
As was numerically studied\cite{FDBV,GKK,DGMS}, 
the maximum eigenvalue for the AB-model is bounded by
$\sqrt{k_{max}}$ such that $\lambda_{max} \le const.\sqrt{k_{max}}$,
where $k_{max}$ means the maximum order of nodes.
And the numerical studies showed that
$k_{max} \propto \sqrt{N}$.
And therefore, the numerical studies validated
$\lambda_{max} \le const.N^{1/4}$.
On this problem, very recently, Chung, Lu and Vu\cite{CLV} have
proved a very general theorem:
In a complex network model, $\lambda_{max}$ is always bounded 
by the lower and upper bounds
such that
$a \le \lambda_{max} \le b$.
Define the second order average degree of nodes, $\tilde{d}$ (see Sec.IV).
Then, (C1) if the exponent $\gamma > 2.5$, then 
$const. \tilde{d} \le \lambda_{max} \le const.\sqrt{k_{max}}$.
(C2) If the exponent $\gamma < 2.5$, then 
$const. \sqrt{k_{max}} \le \lambda_{max} \le const. \tilde{d}$.
(C3) And if the exponent $\gamma = 2.5$, then a transition happens.
From this theorem, the AB-model belongs to the first category
since $\gamma = 3$.
However, in spite of the seemingly important theorem,
because of the lack of other good examples other than the AB-model,
the examples for the other categories have not yet been so well-known so far.

So, the purpose of this paper is to study in much details 
the DSFN model proposed by Barab\'{a}si, Ravasz, and Vicsek\cite{BRV}
in order to  give a good example for the other category of the theorem.
This study will provide a rigorous treatment for the complex network model.

The organization of the paper is the following.
In Sec. II, we will introduce the DSFN model that was first studied by
Barab\'{a}si, Ravasz, and Vicsek\cite{BRV}.
In Sec. III, we will present the exact numbers of nodes
and its degrees.
And we calculate the exact scaling behavior of the nodes.
In Sec. IV, we will calculate the exact number of the
second order average degree $\tilde{d}$, using the 
exact number distribution of nodes and degrees.
In Sec. V, we will introduce the so-called
adjacency matrix in the network theory to the DSFN.
In Sec. VI, we will introduce the eigenvalue problem of the
adjacency matrix.
In Sec. VII, we will present the numerical results
of the spectra of the adjacency matrices for the
DSFNs up to the $n = 7$ generation.
And we will derive the exact numbers of degeneracies in the spectrum.
In sec. VIII, we will present an analytical method that
deduces the exact numbers of the degeneracies
in terms of the irreducible polynomials for the DSFN.
We will also present some conjectures for the polynomials.
In Sec. IX, we will discuss the role of the roots
of the irreducible polynomials.
In Sec. X, we show that there is a hidden symmetry
in the adjacency matrix of the DSFN.
In Sec. XI, we will study the zero modes and its 
index theorem in the DSFN.
In Sec. XII, we will discuss the nature of the maximum
eigenvalue in the spectrum of the DSFN.
We will prove several theorems using some mathematical theorems.
And we will discuss the relationship between the results of our theory
and the Chung, Lu, and Vu's theorem\cite{CLV}.
In Sec. XIII, conclusion will be made.

\section{Deterministic Scale-free network}
Let us introduce the DSFN model invented by Barab\'{a}si, Ravasz, and Vicsek\cite{BRV}.
The development of this network is shown in Fig.1. 
The black and red nodes show the hub and rim nodes.
We call the most connected hub and rim the root and leaf, respectively.

\begin{figure}[h]
\includegraphics[scale=0.5]{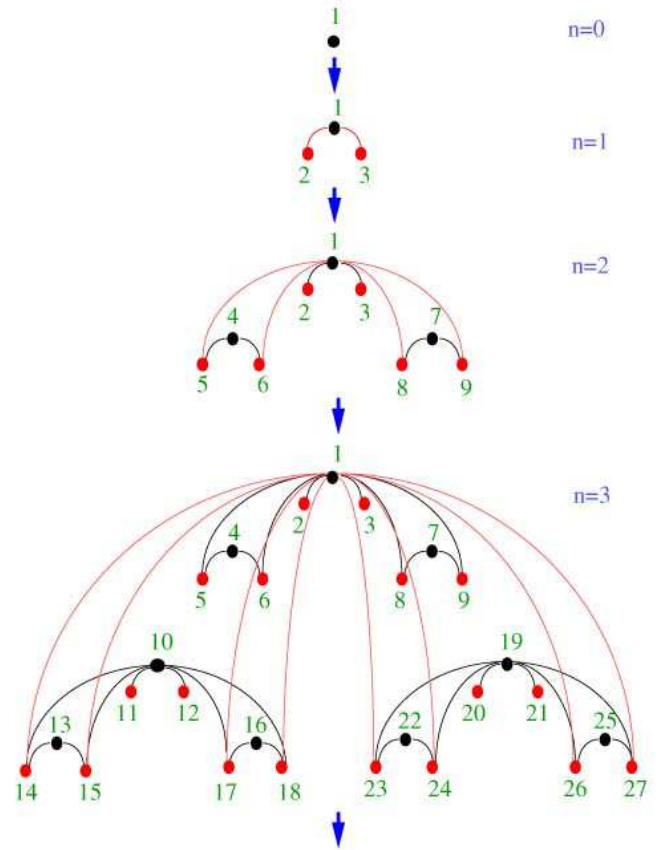}
\caption{
(Color online)
The deterministic scale-free network.
The black and red nodes show the hub and rim nodes.
We call the most connected hub and rim the root and leaf, respectively.
This network is a bipartite structure.
} 
\end{figure}

In this model the total number of nodes, $N(n)$, 
the total number of links, $L(n)$, and
the maximum number of links,  $k_{max}(n)$
are given by
$$N(n) = 3^{n}, \eqno{(5)}$$
$$L(n) = 3L(n-1) + 2^{n} = 2(3^{n} - 2^{n}), \eqno{(6)}$$
$$k_{max}(n) = 2 + 2^{2} + \cdots + 2^{n} = 2(2^{n}-1). \eqno{(7)}$$
respectively. 
From these we find the their development as follows:
As $n = 0, 1, 2, 3, 4, \cdots$, then
$$N(n) = 1, 3, 9, 27, 81, \cdots, \eqno{(8)}$$
$$L(n) = 0, 2, 10, 38, 130, \cdots, \eqno{(9)}$$
$$k_{max}(n) = 0, 2, 6, 14, 30, \cdots. \eqno{(10)}$$

Let us consider the average link number (i.e., the average degree) $\langle k \rangle$ of a network.
It is defined by
$$\langle k \rangle(n) \equiv \frac{1}{N(n)}\sum_{i=1}^{N(n)} k_{i}.    \eqno{(11)}$$
The meaning of this is just the number of links per node  (i.e., the average degree).
We may call it the first order average degree.
On the other hand, 
if we use the number $P(k)$ of nodes with $k$ degree, 
then we can write the average as
$$\langle k \rangle(n) =\frac{1}{N(n)}\sum_{k=1}^{k_{max}(n)} kP(k).    \eqno{(12)}$$
Therefore, we find that the conversion is carried out by
$$P(k) = \sum_{i=1}^{N(n)} \delta_{k,k_{i}}.                       \eqno{(13)}$$ 
Now, we are able to calculate $\langle k \rangle$ for our DSFN.
Substituting Eq.s (8) and (9) into Eq.(11),
we obtain
$$\langle k \rangle(n) =\frac{1}{N(n)}\sum_{i=1}^{N(n)} k_{i} = \frac{2L(n)}{N(n)}$$
$$= \frac{2\cdot 2(3^{n} - 2^{n})}{3^{n}}=4\left(1-\left( \frac{2}{3}\right)^{n}\right)
\stackrel{n \rightarrow \infty} {\rightarrow } 4.                   \eqno{(14)}$$

In this way, even though the network becomes very complex as $n \rightarrow \infty$,
the average approaches a finite constant $4$.
This is due to the following fact:
In this DSFN as the iteration is repeated,
the order of the most connected hub becomes large indefinitely
while its number remains very few (i.e., $\approx o(1)$).  
On the other hand,
the numbers of the very few connected nodes become large indefinitely.
Hence the sum of the magnitude of order ($k$) times the number ($P(k)$) of nodes with the order $k$ 
may remain finite.

\section{Exact numbers of nodes and degrees}
Let us find the exact numbers of nodes and degrees.
This would be very crucial for our later purposes
in order to evaluate many quantities in the network theory.
 
As was discussed by Barab\'{a}si, Ravasz, and Vicsek\cite{BRV},
in the DSFN there are two categories of nodes called "hub" nodes and the "rim" nodes.
As they called the most connected hub node the "root" (shown as black dots in Fig.1),
we would like to call the most connected rim node the "leaf" (shown as red dots in Fig.1).
From seeing Fig.1, the locations of the root node and the leaf nodes
look very similar to those of a hub and rims in an umbrella.
While there exists only one roof node in each generation
of the network, the number of the leaves can increase very rapidly.

Let us first consider the hub nodes.   
In the $i$-th step, the degree of the root, is $2^{i+1}-2$.   
In the next iteration two copies of this hub will appear in the two newly added units.  
As we iterate further, in the $n$-th step $3^{n-i}$ copies of this hub will be present in the network. 
However, the two newly created copies will not increase their degree after further iterations.  
Therefore, after $n$ iterations there are $2\cdot 3^{n-i-1}$ nodes with degree $2^{i+1}-2$.

Let us next consider the rim nodes.   
In the $i$-th step, the degree of the most connected rim, the leaf, is $i$.  
And the number of the such nodes is $2^{i}$.    
In the next iteration one copy of the leaves will be kept the same 
and two copies of the leaves will appear in the two newly added.  
As I iterate further, in the $n$-th step $3^{n-i}$ copies of the leaves will be present in the network. 
Therefore, after $n$ iterations there are $2^{i}\cdot 3^{n-i-1}$ nodes with degree $i$.

Now, denote by $k$ the degree of nodes and 
denote by $P(k)$ the total number of nodes with degree $k$. 
Hence we get the following:\\

\begin{tabular}{|c|c|c|c|}  \hline \hline
\multicolumn{2}{|c|}{\em root nodes}&
\multicolumn{2}{|c|}{\em leaf nodes}\\  \hline 

$k$&$P(k)$&$k$&$P(k)$\\ \hline
$2$&$2\cdot 3^{n-2}$&$1$&$2\cdot 3^{n-2}$\\   \hline
$6$&$2\cdot 3^{n-3}$&$2$&$2^{2}\cdot 3^{n-3}$\\   \hline
$14$&$2\cdot 3^{n-4}$&$3$&$2^{3}\cdot 3^{n-4}$\\   \hline
$\vdots$&$\vdots$&$\vdots$&$\vdots$\\   \hline
$2^{i+1}-2$&$2\cdot 3^{n-1-i}$&$i$&$2^{i}\cdot 3^{n-1-i}$\\   \hline
$\vdots$&$\vdots$&$\vdots$&$\vdots$\\   \hline
$2^{n-1}-2$&$2\cdot 3$&$n-2$&$2^{n-2}\cdot 3$\\   \hline
$2^{n}-2$&$2$&$n-1$&$2^{n-1}\cdot 1$\\   \hline
$2^{n+1}-2$&$1$&$n$&$2^{n}$\\   \hline\hline
\end{tabular}\\
\\ \\
TABLE 1.  The number $P(k)$ of nodes with degree $k$ for the root nodes and leaf nodes.
\\

As was shown by Barab\'{a}si, Ravasz, and Vicsek\cite{BRV}, 
consideration for the root nodes is enough to derive the scaling exponent
of the distribution function $P(k)$ for the root nodes in the network.  
Picking up the $2\cdot 3^{n-1-i}$ nodes with degree $2^{i+1}-2$, 
we can regard $P(k)$ as $2\cdot 3^{n-1-i}$ and $k$ as $2^{i+1}-2$.    
Then eliminating $i$, we can derive
$$P(k) \propto k^{-\gamma}, \eqno{(15)}$$
where 
$$\gamma = \frac{\ln 3}{\ln 2} = 1.584962.  \eqno{(16)}$$
This shows a scale-free nature (i.e., the fractal nature) of the
hub nodes in the network as we expect\cite{Note1}.

On the other hand, it is not true for the leaves in the network.
In this case, regarding $P(k)$ as $2^{i}\cdot 3^{n-1-i}$ and $k$ as $i$,
we find
$$P(k) \propto\left( \frac{2}{3}\right)^{k} = e^{-\gamma' k}, \eqno{(17)}$$
where 
$$\gamma = \ln \left(\frac{3}{2}\right) = 0.405465.  \eqno{(18)}$$
This shows that the scaling nature of the leaf nodes is not scale-free
but exponential.

In this way, the scaling nature of the roots and that of the leaves 
in the DSFN are different from each other.
Thus, we are led to a a certain model which consists of a {\it multifractal} nature of the
complex networks.

\section{The second order average degree}
Let us calculate the second order average degree, $\tilde{d}(n)$.
It is defined by
$$\tilde{d}(n) \equiv \frac{1}{L(n)}\sum_{i=1}^{N(n)} k_{i}^{2}
=\frac{1}{L(n)}\sum_{k=1}^{k_{max}(n)} k^{2}P(k).                  \eqno{(19)}$$
This quantity was recently introduced by Chung, Lu and Vu\cite{CLV}.
Roughly speaking, the meaning of this quality is the average degree per link.
In other words, it is the average degree weighted with the preferential attachment such that
$$\tilde{d}(n) \equiv\sum_{i=1}^{N(n)}  k_{i} \Pi_{i}(k_{i}).      \eqno{(20)}$$
From Eq.(14) we have 
$$\langle k \rangle(n) =\frac{1}{N(n)}\sum_{i=1}^{N(n)} k_{i} = \frac{2L(n)}{N(n)},\eqno{(21)}$$
we derive
$$\tilde{d}(n) = 2\frac{\langle k^{2} \rangle(n)}{\langle k \rangle(n)},    \eqno{(22)}$$
where $\langle k^{2} \rangle(n)$ is defined by
$$\langle k^{2} \rangle(n) =\frac{1}{N(n)}\sum_{i=1}^{N(n)} k_{i}^{2}
 =\frac{1}{N(n)}\sum_{k=1}^{k_{max}(n)} k^{2}P(k),                 \eqno{(23)}$$
the second moment per node.
As was shown before, the average degree $\langle k \rangle(n)$ converges to $4$ 
as $n \rightarrow \infty$.
Hence, the second order average degree $\tilde{d}(n)$ becomes proportional to 
the second moment $\langle k^{2} \rangle(n)$ in the limit
such that
$$\tilde{d} = \frac{\langle k^{2} \rangle(n)}{2}.       \eqno{(24)}$$

Before going to calculate the second order average degree, 
let us first check whether or not the distributions given in the tables
reproduce the correct results for the total numbers of nodes and links.
This problem is trivial.
But as we will see, this is very instructive for our purpose here.

Let us show that the distributions of the nodes and degrees are exact.
Let us first calculate the total number of nodes, once again.
To do so, we just sum them up as follows.
For the root nodes we obtain the following:
$$\sum_{k\in root}P(k) = 1 + 2 + 2\cdot 3 + \cdots + 2\cdot 3^{n-2}$$
$$= 1+2(1 + 3 + \cdots + 3^{n-2}) 
= 1 + 2\left(\frac{3^{n-1}-1}{3-1} \right) = 3^{n-1}.       \eqno{(25)}$$
This is the total number of the root (hub) nodes (i.e., the total number of black dots in Fig.1).
For the leaf nodes we have the following:
$$\sum_{k\in leaf}P(k) = 2\cdot 3^{n-2} + 2^{2}\cdot 3^{n-3} + \cdots + 2^{n-2}\cdot 3 + 2^{n-1}\cdot 1 + 2^{n}$$
$$= 3^{n-1} + 2\cdot 3^{n-2} + 2^{2}\cdot 3^{n-3} + \cdots + 2^{n-2}\cdot 3 + 2^{n-1}\cdot 1 + 2^{n} - 3^{n-1}$$
$$= \left(\frac{3^{n}-2^{n}}{3-2} \right) + 2^{n} - 3^{n-1} = 3^{n}-3^{n-1} = 2\cdot 3^{n-1}.       \eqno{(26)}$$
This is the total number of the leaf (rim) nodes (i.e., the total number of red dots in Fig.1).
Hence, by the addition of both we obtain the desired result:
$$\sum_{k\in all}P(k) = 3^{n} = N(n).                    \eqno{(27)}$$
This proves Eq.(5).

Let us next consider the total number of links in the network.
We calculate it for the root and leaf nodes separately.
For the root nodes, we have the following:
$$\sum_{k \in root}kP(k) = (2^{n+1}-2)\cdot 1 + (2^{n}-2)\cdot 2 + (2^{n-1}-2)\cdot 2\cdot 3$$
$$+ \cdots + (2^{3}-2)\cdot 2\cdot 3^{n-3}  + (2^{2}-2)\cdot 2\cdot 3^{n-2}$$
$$= 2^{n+1}\cdot 1 + 2^{n}\cdot 2 +  2^{n-1}\cdot 2 \cdot 3
+ \cdots + 2^{3}\cdot 2 \cdot 3^{n-3}  + 2^{2}\cdot 2 \cdot 3^{n-2}$$
$$-\left(2 + 2^{2} + 2^{2}\cdot3
+ \cdots + 2^{2}\cdot3^{n-3} + 2^{2}\cdot3^{n-2} \right)$$
$$= 2^{n+1} - 2 + 2^{3}(2^{n-2} + 3\cdot 2^{n-3} + \cdots + 3^{n-3}\cdot 2 + 3^{n-2})$$
$$- 2^{2}(1+ 3+ \cdots + 3^{n-2})$$
$$= 2^{n+1} - 2 + 2^{3} \left( \frac{3^{n-1}-2^{n-1}}{3-2}\right) 
- 2^{2}\left( \frac{3^{n-1}-1}{3-1}\right)$$ 
$$= 2(3^{n} - 2^{n}).                                              \eqno{(28)}$$
Similarly, for the leaf nodes, we have the following:
$$\sum_{k \in leaf}kP(k) 
= 1\cdot 2\cdot 3^{n-2} + 2\cdot 2^{2}\cdot 3^{n-3} + 3\cdot 2^{3}\cdot 3^{n-4} $$
$$+ \cdots + (n-2) 2^{n-2}\cdot 3 + (n-1) 2^{n-1}\cdot 1 + n2^{n}$$
$$=2(1\cdot 3^{n-2} + 2\cdot 2\cdot 3^{n-3} + 3\cdot 2^{2}\cdot 3^{n-4} 
+ \cdots $$
$$+ (n-2) 2^{n-3}\cdot 3 + (n-1) 2^{n-2}\cdot 1 + n2^{n-1}).   \eqno{(29)}$$
At first glance, the summation of Eq.(29) seems quite tough.
But to perform this calculation we can use a mathematical trick as follows:
Define a function $f(x)$ of $x$ such that
$$f(x) = x\cdot3^{n-2} + x^{2}\cdot3^{n-3} + x^{3}\cdot3^{n-4} + \cdots$$ 
$$+ x^{n-2}\cdot 3 + x^{n-1}\cdot 1 + x^{n}.                     \eqno{(30)}$$
Differentiating this with respect to $x$,
we obtain
$$f'(x) = 3^{n-2} + 2x\cdot3^{n-3} + 3x^{2}\cdot3^{n-4} + \cdots$$ 
$$+ (n-2)x^{n-3}\cdot 3 + (n-1)x^{n-2}\cdot 1 + nx^{n-1}.       \eqno{(31)}$$
Substituting $x=2$ into Eq.(31) and comparing it with Eq.(29), we find
$$\sum_{k \in leaf}kP(k) =  2f'(2).                           \eqno{(32)}$$
Therefore, once $f(x)$ is found as a simple function of $x$, we can calculate the sum.
This can be done as follows.
$$f(x) = x\cdot3^{n-2} + x^{2}\cdot3^{n-3} + x^{3}\cdot3^{n-4} + \cdots $$
$$+ x^{n-2}\cdot 3 + x^{n-1}\cdot 1 + x^{n}$$
$$= x(1\cdot3^{n-2} + x\cdot3^{n-3} + x^{2}\cdot3^{n-4} + \cdots + x^{n-3}\cdot 3 + x^{n-2}\cdot 1) + x^{n}$$
$$= x \left( \frac{3^{n-1} - x^{n-1}}{3-x}\right) + x^{n}.                \eqno{(33)}$$
By differentiating this with respect to $x$,
we obtain
$$f'(x) = \frac{3^{n-1} - x^{n-1}}{3-x}$$
$$+ x \left[ \frac{-(n-1)x^{n-2}}{3-x} + \frac{3^{n-1} - x^{n-1}}{(3-x)^{2}}\right]$$
$$ + nx^{n-1}.                                                             \eqno{(34)}$$
Therefore, substituting $x=2$ into Eq.(34), we obtain
$$f'(2) = \frac{3^{n-1} - 2^{n-1}}{3-2} $$
$$+ 2 \left[ \frac{-(n-1)2^{n-2}}{3-2} + \frac{3^{n-1} - 2^{n-1}}{(3-2)^{2}}\right]$$
$$ + n2^{n-1}$$
$$= (3^{n-1} - 2^{n-1}) + 2 \left[ -(n-1)2^{n-2} + 3^{n-1} - 2^{n-1}\right] + n2^{n-1}$$
$$= 3^{n} - 2^{n}.                                                    \eqno{(35)}$$
Substituting this into Eq.(32), we obtain
$$\sum_{k \in leaf}kP(k) =  2(3^{n} - 2^{n}).                           \eqno{(36)}$$
In this way, explicitly using the exact numbers of nodes and degrees,
we can show that each sum produces 
the total number of links in the network as we expected.
Hence, this proves Eq.(6).
This situation encourages us to perform the calculation of the
second order average degree of Eq.(19).

Let us next do this.
In Eq.(19), we need to separate it into two parts of the sum as follows:
$$\tilde{d}(n) \equiv \frac{1}{L(n)}\sum_{i=1}^{N(n)} k_{i}^{2}
=\frac{1}{L(n)}\sum_{k=1}^{k_{max}(n)} k^{2}P(k)$$
$$= \frac{1}{L(n)}\left( \sum_{k\in root} k^{2}P(k) 
+ \sum_{k\in leaf} k^{2}P(k)\right).                  
                                                              \eqno{(37)}$$
Let us consider the sum for the root nodes.
Using the previous result, we obtain the following:
$$\sum_{k\in root} k^{2}P(k) = (2^{n+1}-2)^{2}\cdot 1  
+ (2^{n}-2)^{2}\cdot 2 + (2^{n-1}-2)^{2}\cdot 2\cdot 3$$
$$+ \cdots + (2^{3}-2)^{2}\cdot 2\cdot 3^{n-3}  + (2^{2}-2)^{2}\cdot 2\cdot 3^{n-2}$$
$$= (2^{n+1}-2)^{2}$$
$$+2[(2^{2})^{n}+(2^{2})^{n-1}3+ \cdots + (2^{2})^{2}3^{n-2}]$$
$$-8[2^{n}+2^{n-1}3+ \cdots + 2^{2}3^{n-2}]$$
$$+8[1+3+\cdots +3^{n-2}]$$
$$= (2^{n+1}-2)^{2}$$
$$+2^{5}[(2^{2})^{n-2}+(2^{2})^{n-3}3+ \cdots + 2^{2}3^{n-3}+ 3^{n-2}]$$
$$-2^{5}[2^{n-2}+2^{n-3}\cdot 3+ \cdots + 2\cdot 3^{n-3} + 3^{n-2}]$$
$$+2^{3}[1+3+\cdots +3^{n-2}]$$
$$= (2^{n+1}-2)^{2} +2^{5}\frac{(2^{2})^{n-1}-3^{n-1}}{2^{2}-3}$$
$$ -2^{5}\frac{3^{n-1}-2^{n-1}}{3-2}
+ 2^{3}\frac{3^{n-1}-1}{3-1}$$
$$= (2^{n+1}-2)^{2} + 2^{5}(4^{n-1}-2\cdot 3^{n-1}+2^{n-1})
+ 2^{2}(3^{n-1}-1).                                            \eqno{(38)}$$

Let us consider the sum for the leaf nodes.
$$\sum_{k \in leaf}k^{2}P(k)$$ 
$$= 1^{2}\cdot 2\cdot 3^{n-2} + 2^{2}\cdot 2^{2}\cdot 3^{n-3} 
+ 3^{2}\cdot 2^{3}\cdot 3^{n-4} + \cdots $$
$$+ (n-2)^{2} 2^{n-2}\cdot 3 + (n-1)^{2} 2^{n-1}\cdot 1 + n^{2}2^{n}$$
$$=2(1^{2}\cdot 3^{n-2} + 2^{2}\cdot 2\cdot 3^{n-3} 
+ 3^{2}\cdot 2^{2}\cdot 3^{n-4} + \cdots$$
$$+ (n-2)^{2} 2^{n-3}\cdot 3 + (n-1)^{2} 2^{n-2}\cdot 1 + n^{2}2^{n-1}).   \eqno{(39)}$$
To calculate Eq.(39) we need a similar trick as before.
Using Eq. (31), we can define
$$xf'(x) = x3^{n-2} + 2x^{2}\cdot3^{n-3}
 + 3x^{3}\cdot3^{n-4} + \cdots$$ 
$$+ (n-2)x^{n-2}\cdot 3 + (n-1)x^{n-1}\cdot 1 + nx^{n}.                      \eqno{(40)}$$
Differentiating this with respect to $x$ yields
$$(xf'(x))' = 3^{n-2} + 2^{2}x\cdot3^{n-3} 
+ 3^{2}x^{2}\cdot3^{n-4} + \cdots $$
$$+ (n-2)^{2}x^{n-3}\cdot 3 + (n-1)^{2}x^{n-2}\cdot 1 + n^{2}x^{n-1}.       \eqno{(41)}$$
Therefore, substituting $x=2$ into this, we get
$$(xf'(x))'\mid_{x=2} = 3^{n-2} + 2^{2}\cdot 2\cdot 3^{n-3} 
+ 3^{2}\cdot 2^{2}\cdot3^{n-4} + \cdots$$ 
$$+ (n-2)^{2}2^{n-3}\cdot 3 + (n-1)^{2}2^{n-2}\cdot 1 + n^{2}2^{n-1}.       \eqno{(42)}$$
Comparing this with Eq.(39), we get the relation
$$\sum_{k \in leaf}k^{2}P(k) = 2 (xf'(x))'\mid_{x=2}.                       \eqno{(43)}$$

Let us evaluate this. 
Using Eq.(34), we obtain
$$xf'(x) = x\left(\frac{3^{n-1} - x^{n-1}}{3-x}\right)$$
$$+ x^{2} \left[ \frac{-(n-1)x^{n-2}}{3-x} 
+ \frac{3^{n-1} - x^{n-1}}{(3-x)^{2}}\right]$$
$$ + nx^{n}.                                                                  \eqno{(44)}$$
Differentiating this with respect to $x$, we have
$$(xf'(x))' = \frac{3^{n-1} - x^{n-1}}{3-x}$$
$$+ 3x \left[ \frac{-(n-1)x^{n-2}}{3-x} 
+ \frac{3^{n-1} - x^{n-1}}{(3-x)^{2}}\right]$$
$$+ x^{2} \left[ \frac{-(n-1)(n-2)x^{n-3}}{3-x} 
+ \frac{-2(n-1)x^{n-2}}{(3-x)^{2}}\right]$$
$$+x^{2} \left[2 \frac{3^{n-1} - x^{n-1}}{(3-x)^{3}}\right]+ n^{2}x^{n-1}.       \eqno{(45)}$$
Substituting $x = 2$ into this, we get
$$(xf'(x))'\mid_{x=2} = \frac{3^{n-1} - 2^{n-1}}{3-2}$$
$$+ 3\cdot 2 \left[ \frac{-(n-1)2^{n-2}}{3-2} 
+ \frac{3^{n-1} - 2^{n-1}}{(3-2)^{2}}\right]$$
$$+ 2^{2} \left[ \frac{-(n-1)(n-2)2^{n-3}}{3-2} 
+ \frac{-2(n-1)2^{n-2}}{(3-2)^{2}}\right]$$
$$+ 2^{2} \left[2 \frac{3^{n-1} - 2^{n-1}}{(3-2)^{3}}\right]+ n^{2}2^{n-1}$$
$$=3^{n-1}-2^{n-1}+6[-(n-1)2^{n-2}+3^{n-1} - 2^{n-1}] $$
$$+2^{2}[-(n-1)(n-2)2^{n-3}-2(n-1)2^{n-2} + 2(3^{n-1} - 2^{n-1})]$$
$$ + n^{2}2^{n-1}$$
$$=3^{n-1}-2^{n-1} -6(n-1)2^{n-1}+6(3^{n-1} - 2^{n-1})$$
$$-(n-1)(n-2)2^{n-1}-2(n-1)2^{n} + 2^{3}(3^{n-1}-2^{n-1}) + n^{2}2^{n-1}$$
$$=(1+2+2^{2}+2^{3})(3^{n-1}-2^{n-1}) -3(n-1)2^{n}$$
$$ +\{n^{2}-(n-1)(n-2)-4(n-1)\}2^{n-1}$$
$$= 5\cdot 3^{n} - (2n+5)2^{n}.                                \eqno{(46)}$$
Substituting this into Eq.(43), we get
$$\sum_{k \in leaf}k^{2}P(k) = 2(5\cdot 3^{n} - (2n+5)2^{n}).   \eqno{(47)}$$
Using Eq.(47) together with Eq.(38), we obtain 
$$\sum_{k \in all}k^{2}P(k) 
= (2^{n+1}-2)^{2} + 2^{5}(4^{n-1}-2\cdot 3^{n-1}+2^{n-1})$$
$$+ 2^{2}(3^{n-1}-1) + 2(5\cdot 3^{n} - (2n+5)2^{n}).            \eqno{(48)}$$
Therefore, the above procedure enables us to evaluate 
the second order average degree as follows:
$$\tilde{d}(n) \equiv \frac{1}{2L(n)} \sum_{k\in all} k^{2}P(k)$$
$$= \frac{1}{4(3^{n}-2^{n})}[ 
(2^{n+1}-2)^{2} + 2^{5}(4^{n-1}-2\cdot 3^{n-1}+2^{n-1})$$
$$+ 2^{2}(3^{n-1}-1) + 2(5\cdot 3^{n} - (2n+5)2^{n})]$$
$$=\frac{1}{1-\left(\frac{2}{3}\right)^{n}}\left[3\left( \frac{4}{3}\right)^{n} 
- \left(n +\frac{1}{2}\right)\left(\frac{2}{3}\right)^{n} -\frac{5}{2}\right]$$
$$\stackrel{n \rightarrow \infty} {\rightarrow} 3\left( \frac{4}{3}\right)^{n}
 \rightarrow \infty.                                                       \eqno{(49)}$$
In this way, the second order average degree $\tilde{d}$ is calculated explicitly.
We are able to show that it diverges as an exponetial-law.

\section{Adjacency matrix}
Let us consider the adjacency matrix, $A$, 
in the theory of network\cite{FFF,Barabasi,BRV,RB,CLV}.
This matrix is very important for the theory of network,
since it represents the topology of the network structure.
Denote by $a_{ij}$ the component corresponding to the link 
between the $i$-th node and the $j$-th node.
In the theory of network, the elements have only $0$ or $1$
according to whether or not there is a link.
Therefore, in general, the adjacency matrix is defined by

$$a_{ij} = \left\{
\begin{array}{cc}
1 & \mbox{if there is a link}\\
0 & \mbox{otherwise}
\end{array}\right..                        \eqno{(50)}$$

Let us consider the adjacency matrix for the DSFN.
Denote by $A_{n}$ the adjacency matrix for the $n$-th generation of the network.
Using the numbering given in Fig.1, the adjacency matrix is defined
by the following.
$$A_{1} =  \left(
\begin{array}{ccc}
0 & 1 & 1 \\
1 & 0 & 0 \\
1 & 0 & 0 \\
\end{array} 
\right),
                          \eqno{(51)}$$
$$A_{2} =  \left(
\begin{array}{ccccccccc}
0 & 1 & 1 & 0 & 1 & 1 & 0 & 1 & 1\\
1 & 0 & 0 &   &   &   &   &   &  \\
1 & 0 & 0 &   &   &   &   &   &  \\
0 &   &   & 0 & 1 & 1 &   &   &  \\
1 &   &   & 1 & 0 & 0 &   &   &  \\
1 &   &   & 1 & 0 & 0 &   &   &  \\
0 &   &   &   &   &   & 0 & 1 & 1\\
1 &   &   &   &   &   & 1 & 0 & 0\\
1 &   &   &   &   &   & 1 & 0 & 0
\end{array} 
\right),                          \eqno{(52)}$$
and so forth.
Here all other blanks stand for zeros.
We omit them just for seeing the sparse matrix structure corresponding to
the network geometry of the DSFN.
What is important here in the above is that the adjacency matrix
for a certain generation of the network can be
almost 3-block diagonal by those for the last generation of the network.
From this, the fractal nature of the adjacency matrices for the DSFN is now very clear.

\section{Eigenequations}
Let us consider the eigenvalue problem for the adjacency matrix in the DSFN.
The eigenequation for the $n$-th generation of the network is given by
$$A_{n}\vec{X}_{n} = \lambda \vec{X}_{n},             \eqno{(53)}$$
$$\vec{X}_{n}^{t}\cdot \vec{X}_{n}  = 1,                        \eqno{(54)}$$
where 
$A_{n}$ is the $3^{n}\times 3^{n}$ matrix and 
$\vec{x}_{n}$ the $3^{n}$-dimensional vector
with its transpose $\vec{x}_{n}^{t}$.
This reduces to the following eigenvalue problem:
$$\det[\lambda I_{n} -A_{n}] = 0,                            \eqno{(55)}$$
where $I_{n}$ stands for the $3^{n}\times 3^{n}$ unit matrix.
Denote by $D_{n}(\lambda)$ the determinant 
in the left hand side such that
$$D_{n}(\lambda) =  \det[\lambda I_{n} -A_{n}].                 \eqno{(56)}$$
This can be formally expanded with respect to $\lambda$ as
$$D_{n}(\lambda) = \lambda^{3^{n}} - a_{1}\lambda^{3^{n}-1} 
+ a_{2}\lambda^{3^{n}-2} + \cdots $$
$$+ (-1)^{3^{n}-1}a_{3^{n}-1}\lambda + (-1)^{3^{n}} a_{3^{n}}.  \eqno{(57)}$$

We note the following:
Since the $\lambda$ is an eigenvalue of the adjacency matrix $A_{n}$, 
from the knowledge of linear algebra, we are able to derive the following:
$$D_{n}(A_{n}) = A_{n}^{3^{n}} - a_{1}A_{n}^{3^{n}-1} + a_{2}A_{n}^{3^{n}-2}
+ \cdots$$
$$ + (-1)^{3^{n}-1}a_{3^{n}-1}A_{n} + (-1)^{3^{n}} a_{3^{n}}I_{n} = 0.  \eqno{(58)}$$

For example, we easily have the following:
$$D_{1}(A_{1}) = A_{1}^{3} - a_{1}A_{1}^{2} + a_{2}A_{1} - a_{3}I_{1} = 0,  \eqno{(59)}$$
$$D_{2}(A_{2}) = A_{n2}^{9} - a_{1}A_{2}^{8} + a_{2}A_{2}^{7}
- a_{3}A_{2}^{6} + a_{4}A_{2}^{5}$$
$$ - a_{5}A_{2}^{4} + a_{6}A_{2}^{2}
- a_{7}A_{2} + a_{9}I_{3} = 0,                                         \eqno{(60)}$$
and so forth.
As we will see later, all the terms of even number powers of $\lambda$ vanish in the DSFN.
This is attributed to the topology of the network.

\section{Numerical calculation of the spectra}
Let us obtain the spectra of the adjacency matrices, discussed in the above.
We use a computer calculation for this purpose.
The calculated spectra for the adjacency matrix for each generation of the
network are shown in Fig.2.

\begin{figure}[h]
\includegraphics[scale=0.55]{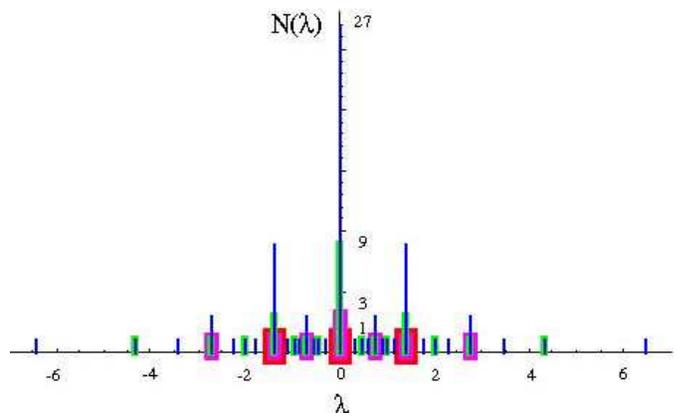}
\caption{
(Color online)
The spectrum of adjacency matrix for the deterministic scale-free network.
The spectrum of adjacency matrix for the network of $n$-th the generation is shown for
$n=1$ (red); $n=2$ (pink); $n=3$ (green); $n=4$ (blue), respectively.} 
\end{figure}

From the numerical results,
we find the following very important characters of the spectra:
(i) The maximum eigenvalue, 
$\lambda_{1}^{(n)}$, 
at the $n$-th generation of the network becomes the second largest eigenvalue, 
$\lambda_{2}^{(n+1)}$, 
at the $(n+1)$-th generation of the network.

(ii) Similarly for the other eigenvalues,
all the eigenvalues appeared at the previous generations of the network
always exist in the eigenvalues appeared at the newly generation of the network.

(iii) {\it The spectrum consists of highly degenerate levels}.
For example, for $n=1$ there are three levels (shown by red) of 
$\lambda = 0, \pm \sqrt{2}$ with single degeneracy.
For $n=2$ there are nine levels (shown by pink) in the spectrum. 
There is only one peak at the center level of $\lambda = 0$ with degeneracy of $3$ 
and the other levels of 
$\lambda = \pm (\sqrt{3}-1), \pm \sqrt{2}, \pm(\sqrt{3}+1)$
are all single levels.
For $n=3$ there are 27 levels (shown by green) in the spectrum.
There is a highest peak at the center level ($\lambda = 0$) with degeneracy of $9$.
There are two peaks at the levels of $\lambda = \pm \sqrt{2}$ with degeneracy of $3$.
And the other 12 levels are all single levels.
For $n=4$ there are 81 levels (shown by blue) in the spectrum.
There is a highest peak at the center level ($\lambda = 0$) with degeneracy of $27$.
There are two peaks at the levels of $\lambda = \pm \sqrt{2}$ with degeneracy of $9$.
There are four peaks at the levels of $\lambda = \pm \sqrt{2}$ with degeneracy of $3$.
And the other 24 levels are all single levels,
and so forth.

In order to study this nature further,
we have done numerical calculations for the spectra
of the adjacency matrices with the sizes of $3^{5}=243, 3^{6}=729, 3^{7}=2187,
3^{8}=6561$, respectively.
From this we have confirmed ourselves that this nature is numerically exact
at any generation of the network up to $n=8$.
We show this in Fig.3 for the cases of
$3^{5}=243$ (green), $6^{6}=729$(blue), $3^{7}=2187 $(red).

\begin{figure}[h]
\includegraphics[scale=0.65]{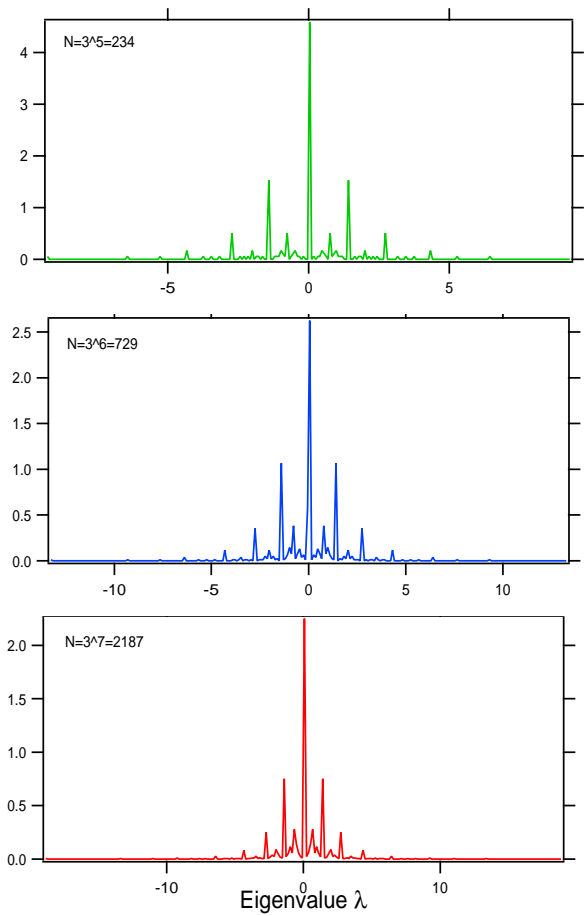}
\caption{
(Color online)
The spectrum of adjacency matrix for the deterministic scale-free network.
The spectrum of adjacency matrix for the network of $n$-th the generation is shown for
$n=5$ (green); $n=6$ (blue); $n=7$ (red), respectively.
} 
\end{figure}

The above nature is remarkable.
It enables us to calculate the exact sequence of the degeneracies in the spectrum
at any generation of the network,
apart from finding the exact eigenvalues.
By counting the numbers of the levels and its degeneracies,
we find the following very important result.

(iv) Let us denote $\mu_{j}$ the degeneracy of the $j$-th peak 
in the spectrum for the network of the $n$-th generation.
Denote by $Q_{j}$ the number of the eigenvalues having the same degeneracy of $\mu_{j}$.
We find 
$$\mu_{j} = 3^{j} \ \ \mbox{for $j=0, 1, \cdots, n-1$},    \eqno{(61)}$$
$$Q_{j} = \left\{
\begin{array}{ll}
2^{n-1-j} & \mbox{for $j=1, 2, \cdots, n-1$}\\
3\cdot 2^{n-1} & \mbox{for $j=0$}
\end{array}
\right..                                          \eqno{(62)}$$

From this we can check whether or not the above formula is correct.
For this purpose, we just reproduce the total number of 
eigenvalues of the adjacency matrix as follows.
$$\sum_{j=0}^{n-1}\mu_{j}Q_{j} = \mu_{0}Q_{0}+\mu_{1}Q_{1}+\cdots+ \mu_{n-1}Q_{n-1}$$
$$= 1\cdot 2^{n-1}\cdot 3+ 3^{n-1}\cdot 1 + 3^{n-2}\cdot 2 + \cdots + 3\cdot 2^{n-2}$$
$$= 2^{n-1}\cdot 3 +3 (3^{n-2}\cdot 1 + 3^{n-3}\cdot 2 
+ \cdots + 3\cdot 2^{n-3} + 1\cdot 2^{n-2})$$
$$= 2^{n-1}\cdot 3 + 3\left(\frac{3^{n-1}-2^{n-1}}{3-2}\right) = 3^{n}.    \eqno{(63)}$$
This is nothing but the total number of eigenvalues in the spectrum.
Hence, the formula is proved.

Here we would like to point out the eigenvectors (i.e., the states)
corresponding to the eigenvalues as well.
Then we find the following very interesting character:

(v) {\it The eigenvectors (i.e., the states) are very localized}.
The states of 
$\lambda = 0$ 
first appeared in the network of the first generation 
are localized only on the pairs of leaf nodes
such as the pair of 2 and 3, the pair of 5 and 6, and
the pair of 8 and 9, etc.
The states of 
$\lambda = \pm \sqrt{2}$ 
first appeared in the network of the first generation 
are localized only on the smallest triples centered at the root hub
such as the triple of 1, 2, and 3, the triple of 4, 5, and 6, and
the triple of 7, 8, and 9, etc.
The states of 
$\lambda = \pm (\sqrt{3} \pm 1)$ 
first appeared in the network of the second generation 
are localized within the subnetworks having the size of that of the second generation,
such as the network of 1-9, etc.
The states of 
$\lambda = \pm \sqrt{2(3\pm\sqrt{3} \pm \sqrt{11\pm6\sqrt{3}})}$ 
first appeared in the network of the third generation 
are localized within the subnetworks having the size of that of the second generation,
such as the network of 1-27, etc.
and so forth.
And the number of such subnetworks gives its degeneracy.

Thus, we find that the larger the eigenvalue the larger the extent of the eigenvector.
Hence, we would like to conclude that the eigenvector with
the maximum eigenvalue (say, $\lambda_{1}$) is most delocalized
while the eigenvalues of $\lambda = 0$ are extremely localized at the leaf nodes in the network.
This nature explains the meaning of the spectrum shown in Fig.s 2 and 3.
Therefore, since we see similar spectra in most SFN models,
we may expect that the same holds true for every SFN.
This is a very interesting point in the study of SFNs.

\section{Simple observations for $D_{n}(\lambda)$}
We now show why so.
We are going to treat analytically the above result in the previous section.
To do this, let us consider some interesting nature of $D_{n}(\lambda)$.
Let us first consider $D_{1}(\lambda)$.
$$D_{1}(\lambda) = \left|
\begin{array}{ccc}
\lambda & -1 & -1 \\
-1 & \lambda & 0 \\
-1 & 0 & \lambda \\
\end{array} 
\right|.                                              \eqno{(64)}$$
We can calculate this and convert it as follows:
$$D_{1}(\lambda) = \left|
\begin{array}{ccc}
\lambda & -1 & -1 \\
-1 & \lambda & 0 \\
-1 & 0 & \lambda \\
\end{array} 
\right|
=\left|
\begin{array}{ccc}
\lambda & -1 & -1 \\
-1 & \lambda & 0 \\
0 &-\lambda & \lambda \\
\end{array} 
\right|$$
$$=\left|
\begin{array}{ccc}
\lambda & -2 & -1 \\
-1 & \lambda & 0 \\
0 & 0 & \lambda \\
\end{array} 
\right|
=\lambda
\left|
\begin{array}{cc}
\lambda & -2\\
-1 & \lambda\\
\end{array} 
\right|
=\lambda(\lambda^{2}-2).                            \eqno{(65)}$$
Obviously, this provides three eigenvalues of
$$\lambda = 0, \pm\sqrt{2}.                        \eqno{(66)}$$
Therefore, let us denote it as
$$D_{1}(\lambda) = \lambda(\lambda^{2}-2) = \lambda f_{1}(\lambda),   \eqno{(67)}$$
where 
$$f_{1}(\lambda) = \lambda^{2} - 2.                            \eqno{(68)}$$

Let us next consider $D_{2}(\lambda)$.
It is given by
$$D_{2}(\lambda) = \left|
\begin{array}{ccccccccc}
\lambda & -1 & -1 & 0 & -1 & -1 & 0 & -1 & -1\\
-1 & \lambda & 0 &   &   &   &   &   &  \\
-1 & 0 &\lambda &   &   &   &   &   &  \\
0 &   &   & \lambda & -1 & -1 &   &   &  \\
-1 &   &   & -1 & \lambda & 0 &   &   &  \\
-1 &   &   & -1 & 0 & \lambda &   &   &  \\
0 &   &   &   &   &   & \lambda & -1 & -1\\
-1 &   &   &   &   &   & -1 & \lambda & 0\\
-1 &   &   &   &   &   & -1 & 0 & \lambda
\end{array} 
\right|,                                        \eqno{(69)}$$
where all blanks stand for zeros.
Doing the same procedure, we find
$$D_{2}(\lambda) = \lambda^{3}\left|
\begin{array}{cccccc}
\lambda & -2 & 0 & 0 & -2 & -2 \\
-1 & \lambda & 0 &   &   &   \\
0 & 0 &\lambda &   &   &  -2 \\
0 &   &   & \lambda & -2 & 0 \\
-1 &   &   & -1 & \lambda & 0 \\
-1 &   & -1 & 0 & 0 & \lambda \\
\end{array} 
\right|$$
$$= \lambda^{3}(\lambda^{2}-2)(\lambda^{4}-8\lambda^{2}+4)$$
$$= \lambda^{3}(\lambda^{2}-2)(\lambda^{2}
-2\lambda-2)(\lambda^{2}+2\lambda-2).                        \eqno{(70)}$$
Hence, we obtain
$$D_{2}(\lambda) = \lambda^{3}f_{1}(\lambda)f_{2}(\lambda),  \eqno{(71)}$$
where
$$f_{2}(\lambda) = \lambda^{4}-8\lambda^{2}+4
=(\lambda^{2}-2\lambda-2)(\lambda^{2}+2\lambda-2)$$
$$\equiv g_{2}(\lambda)h_{2}(\lambda).                        \eqno{(72)}$$
In the same way, we find the following for the determinants
$D_{3}(\lambda)$ and $D_{4}(\lambda)$:
$$D_{3}(\lambda) = \lambda^{9}(\lambda^{2}-2)^{3}
(\lambda^{4}-8\lambda^{2}+4)$$
$$(\lambda^{8} - 24\lambda^{6} + 104\lambda^{4}- 96\lambda^{2} + 16)$$
$$= \lambda^{9}\{f_{1}(\lambda)\}^{3}f_{2}(\lambda)f_{3}(\lambda),   \eqno{(73)}$$
$$D_{4}(\lambda) = \lambda^{27}(\lambda^{2}-2)^{9}
(\lambda^{4}-8\lambda^{2}+4)^{3}$$
$$\times(\lambda^{8} - 24\lambda^{6} + 104\lambda^{4}- 96\lambda^{2} + 16)$$
$$\times(\lambda^{16} - 64\lambda^{14} + 1104\lambda^{12}- 742\lambda^{10} + 22112\lambda^{8}$$
$$-29696\lambda^{6} + 17664\lambda^{4}  -4096\lambda^{2} + 256)$$
$$= \lambda^{27}\{f_{1}(\lambda)\}^{9}\{f_{2}(\lambda)\}^{3}f_{3}(\lambda)f_{4}(\lambda),    \eqno{(74)}$$
and so forth.
Thus, we obtain a series of irreducible polynomials as follows:
$$f_{1}(\lambda) = \lambda^{2} - 2, $$
$$f_{2}(\lambda) = \lambda^{4}-8\lambda^{2}+4 = (\lambda^{2}-2\lambda-2)(\lambda^{2}+2\lambda-2), $$
$$f_{3}(\lambda) = \lambda^{8} - 24\lambda^{6} + 104\lambda^{4}- 96\lambda^{2} + 16, $$
$$f_{4}(\lambda) = \lambda^{16} - 64\lambda^{14} + 1104\lambda^{12}- 742\lambda^{10} + 22112\lambda^{8}$$
$$-29696\lambda^{6} + 17664\lambda^{4}  -4096\lambda^{2} + 256$$
$$=(\lambda^{8} - 4\lambda^{7}-24\lambda^{6} +40\lambda^{5}
+ 104\lambda^{4}-80\lambda^{3}- 96\lambda^{2} + 32\lambda+16)$$
$$\times(\lambda^{8} + 4\lambda^{7}-24\lambda^{6} -40\lambda^{5}
+ 104\lambda^{4}+80\lambda^{3}- 96\lambda^{2} - 32\lambda+16).          \eqno{(75)} $$
Using these polynomials, we are able to represent 
the sequence of the determinants as follows:
$$D_{1}(\lambda) = \lambda f_{1}(\lambda),$$
$$D_{2}(\lambda) = \lambda^{3} f_{1}(\lambda)f_{2}(\lambda),$$
$$D_{3}(\lambda) = \lambda^{9} \{f_{1}(\lambda)\}^{3}f_{2}(\lambda)f_{3}(\lambda),$$
$$D_{4}(\lambda) = \lambda^{27} \{f_{1}(\lambda)\}^{9}
\{f_{2}(\lambda)\}^{3}f_{3}(\lambda)f_{4}(\lambda),                          \eqno{(76)}$$
etc., where
$f_{n}(\lambda)$ is an even function of $2^{n}$-th order polynomial of $\lambda$ such that
$f_{n}(\lambda) = f_{n}(-\lambda)$.
This symmetry in the eigenvalues is due to the bipartite nature of the DSFN.

Therefore, we expect that we can carry out such a factorization of the determinant
at any generation of the network. 
Hence, we are led to state the following conjectures.
\newtheorem{guess}{Conjecture}
\begin{guess}
For the polynomial with even suffix of $n=2m$,
it is always factorized as
$$f_{2m}(\lambda) = g_{m}(\lambda)h_{m}(\lambda),    \eqno{(77)}$$
where $g_{m}(\lambda)$ and $h_{m}(\lambda)$ are
$m$-th order polynomials of $\lambda$.
\end{guess}
\begin{guess}
$$D_{n}(\lambda) = \lambda^{3^{n-1}} \{f_{1}(\lambda)\}^{3^{n-2}}\{f_{2}(\lambda)\}^{^{3^{n-3}}}$$
$$\cdots \{f_{n-2}(\lambda)\}^{3}f_{n-1}(\lambda)f_{n}(\lambda).        \eqno{(78)}$$
\end{guess}

The meaning of the second conjecture is now very clear.
(1) The spectrum is symmetrical around the center level of $\lambda=0$ (see Fig.2 and Fig.3).
This means that if $\lambda$ is an eigenvalue of the adjacency matrix,
then so if for $-\lambda$.
(2) The powers in Eq.(78) represent the numbers of the degeneracies of the eigenvalues.
For example,
the $\lambda=0$ level consists of $3^{n-1}$
(i.e., the degeneracy of $3^{n-1}$)
and this corresponds to the highest peak at the center of $\lambda=0$ in the spectrum.
The $\lambda= \pm\sqrt{2}$ levels consist of $3^{n-2}$
(i.e., the degeneracy of $3^{n-2}$)
and these correspond to the second highest peaks 
at the center of $\lambda=\pm \sqrt{2}$ in the spectrum,
and so forth.
This proves the formulas of Eq.s (61) and (62), previously found.

The validity of the conjectures is also supported by our numerical calculations 
as mentioned before. But the exact proofs have not been made yet, however.

\section{The roots of the irreducible polynomials}
We now consider zeros (i.e., roots) of the irreducible polynomials of $f_{n}(\lambda)$ for
$k=1, 2, \cdots ,n$.
As studied before, the order of $f_{n}(\lambda)$ is $2^{n}$ and
it is a function of $\lambda^{2}$.
Let us denote as $y = \lambda^{2}$.
Then, $f_{n}(\lambda)$ becomes a function of $y$ such as $f_{n}(y)$.
Now we have the following:

$$f_{1}(y) = y - 2, $$
$$f_{2}(y) = y^{2}-8y +4, $$
$$f_{3}(y) = y^{4} - 24y^{3} + 104y^{3}- 96y + 16, $$
$$f_{4}(y) = y^{8} - 64y^{7} + 1104y^{6}- 742y^{5} + 22112y^{4}$$
$$-29696y^{3} + 17664y^{2}  -4096y + 256,                   \eqno{(79)} $$
and so forth.

Since $f_{n}(y)$ is now a $2^{n}$-th order polynomial of $y$,
it has to consist of $2^{n}$ zeros.
Then the meaning of irreducibility is the following:
$f_{n}(y)$ ($k = 1, 2, \cdots, n$) does not share common roots 
with the other generations of the polynomials.
This can be proved by the Sturm theorem for polynomials.\cite{Lang}.

From the knowledge of algebra, if there is a series of the irreducible polynomials,
then the roots of the polynomial of order $n-1$ always exist in the intervals
between the roots of the polynomial of order $n$.
Therefore, the maximum root of the polynomial of order $n$ exceeds
that of order $n-1$.
In our problem, the irreducible polynomial $f_{n}(\lambda)$ is of order $2^{n}$
and it gives the newly appearing eigenvalues.
Therefore, there are $2^{n}$ eigenvalues in the network of the $n$-th generation.

On the other hand, the previously appeared eigenvalues are given as the roots of the
irreducible polynomials of order up to $n-1$,
i.e.,  $f_{1}(\lambda), f_{2}(\lambda),\dots, f_{n-1}(\lambda)$.
Therefore, there are 
$1+2+\cdots +2^{n-1}= 2^{n} -1 $ eigenvalues already in the spectrum.
Thus, the number of newly appearing eigenvalues is exactly one 
more than that of the previously existed eigenvalues.
Hence, by the knowledge of algebra, 
the eigenvalues of the network of the $n$-th generation
sandwiches the previously existed eigenvalues such that
the maximum eigenvalue may exceed
that of the previous generation.
As we iterate the network, $n$ becomes large indefinitely.
Therefore, the maximum eigenvalue develops further.

To understand this point further, we have numerically investigated the
ratio $R(n)$ between the maximum eigenvalue of $\lambda_{max}(n)$ at the $n$-th
generation and that of $\lambda_{max}(n-1)$ at the $(n-1)$-th generation
up to $n = 8$. 
Denote the ratio by $R(n) = \lambda_{max}(n)/\lambda_{max}(n-1)$.  
The result is as follows:
$R(1) = 1.932\cdots,
R(2) = 1.583\cdots,
R(3) = 1.486\cdots,
R(4) = 1.447\cdots,
R(5) = 1.430\cdots,
R(6) = 1.422\cdots,
R(7) = 1.418\cdots.$
From this we see that as $n$ becomes large the ratio tends to
the number $\sqrt{2} = 1.4142\cdots$.
Thus, we are led to the following conjecture:
\begin{guess}
As $n \rightarrow \infty$,       
$$\lambda_{max} \rightarrow 2^{n/2}.      \eqno{(80)}$$
\end{guess}
This conjecture can be proved from the nature of the series of the
irreducible polynomials in Eq.(75).
Let us first consider the case of $n=$ even.
As we have discussed Conjecture 1,
if $n =$ even, then the polynomial can be factorized as
$$y^{2^{n-1}} - \alpha_{1}(n)y^{2^{n-1}-1} + \alpha_{2}(n)y^{2^{n-1}-2} + \cdots$$
$$= \left(\lambda^{2^{n-1}} - \beta_{1}(n)\lambda^{2^{(n-1)} -1} + \cdots\right)$$ 
$$\times \left(\lambda^{2^{n-1}} 
+ \beta_{1}(n)\lambda^{2^{(n-1)} -1} + \cdots\right) = 0.   \eqno{(81)}$$
where
$$\beta_{1}(n) = 2^{n/2}.                                      \eqno{(82)}$$
Therefore, the maximum eigenvalue is given by  
$$\lambda^{2^{n-1}} - \beta_{1}(n)\lambda^{2^{(n-1)} -1} 
- \cdots = 0.                                                  \eqno{(83)}$$
Dividing the above polynomial by $\lambda^{2^{(n-1)} -1}$ yields
$$\lambda = \beta_{1}(n) 
+ O\left(\frac{1}{\lambda}\right) + \cdots.   \eqno{(84)}$$
Since $\lambda_{max} \rightarrow \infty$ as $n \rightarrow \infty$,
we can obtain the approximation of the maximum 
eigenvalue by a perturbation
method.
Hence, for $n =$ even, we obtain
$$\lambda_{max}(n) = 2^{n/2} +  O(2^{(n-1)/2}) + \cdots,          \eqno{(85)}$$ 
such that $\lambda_{max}(n) > 2^{n/2}$.
Similarly, if $n =$ odd, then we can apply the same perturbational 
argument to the irreducible polynomial, then we have 
$$y = a_{1}(n) - O\left(\frac{1}{y}\right) + \cdots,         \eqno{(86)}$$
where $a_{1}(n)= 3\cdot 2^{n}$ and $y = \lambda^{2}$.
Hence, we obtain
$$\lambda_{max}(n) = \sqrt{3}\cdot 2^{n/2}
- O\left(\frac{1}{\lambda}\right) + \cdots,                  \eqno{(87)}$$  
such that $\lambda_{max}(n) < \sqrt{3}\cdot 2^{n/2}$.
This argument supports the conjecture.
Thus, we expect that as $n \rightarrow \infty$
$\lambda_{max}(n) \rightarrow 2^{n/2}$.

In this way, the solution of the series of the irreducible polynomials
is very crucial in finding the spectrum of the DSFN.
This point is supported by our numerical calculations before.

\section{Hidden symmetry in the model}
Let us consider a particular nature of the DSFN.
There is a hidden symmetry in the adjacency matrix, $A_{n}$.
To see this point let us consider the case of $n = 2$.
As was discussed before, the adjacency matrix for this case is given by Eq.(52)
and the eigenvalue problem is 
$$\left(
\begin{array}{ccccccccc}
0 & 1 & 1 & 0 & 1 & 1 & 0 & 1 & 1\\
1 & 0 & 0 &   &   &   &   &   &  \\
1 & 0 & 0 &   &   &   &   &   &  \\
0 &   &   & 0 & 1 & 1 &   &   &  \\
1 &   &   & 1 & 0 & 0 &   &   &  \\
1 &   &   & 1 & 0 & 0 &   &   &  \\
0 &   &   &   &   &   & 0 & 1 & 1\\
1 &   &   &   &   &   & 1 & 0 & 0\\
1 &   &   &   &   &   & 1 & 0 & 0
\end{array} 
\right)
\left(
\begin{array}{c}
x_{1} \\
x_{2} \\
x_{3} \\
x_{4} \\
x_{5} \\
x_{6} \\
x_{7} \\
x_{8} \\
x_{9} 
\end{array} 
\right)
=\lambda
\left(
\begin{array}{c}
x_{1} \\
x_{2} \\
x_{3} \\
x_{4} \\
x_{5} \\
x_{6} \\
x_{7} \\
x_{8} \\
x_{9} 
\end{array} 
\right).                                                    \eqno{(88)}$$
This expression depends on the choice of the arrangement 
of components of the eigenvector
$\vec{X}_{2} = (x_{1}, x_{2}, x_{3},x_{4},x_{5},x_{6},x_{7},x_{8}, x_{9})^{t},$
where $^{t}$ means its transpose.
Reminding ourselves of the bipartite nature of the DSFN,
we can rearrange them as follows:
$$\vec{Y}_{2} = (x_{1}, x_{4}, x_{7},x_{2},x_{5},x_{8},x_{3},x_{6}, x_{9})^{t}. \eqno{(89)}$$
This means that we align the vector-components numbering in $\bmod 3$.
Then we find the following:
$$\left(
\begin{array}{ccccccccc}
  &   &   & 1 & 1 & 1 & 1 & 1 & 1\\
  &   &   & 0 & 1 & 0 & 0 & 1 & 0\\
  &   &   & 0 & 0 & 1 & 0 & 0 & 1\\
1 & 0 & 0 &   &   &   &   &   &  \\
1 & 1 & 0 &   &   &   &   &   &  \\
1 & 0 & 1 &   &   &   &   &   &  \\
1 & 0 & 0 &   &   &   &   &   &  \\
1 & 1 & 0 &   &   &   &   &   &  \\
1 & 0 & 1 &   &   &   &   &   &  
\end{array} 
\right)
\left(
\begin{array}{c}
x_{1} \\
x_{4} \\
x_{7} \\
x_{2} \\
x_{5} \\
x_{8} \\
x_{3} \\
x_{6} \\
x_{9} 
\end{array} 
\right)
=\lambda
\left(
\begin{array}{c}
x_{1} \\
x_{4} \\
x_{7} \\
x_{2} \\
x_{5} \\
x_{8} \\
x_{3} \\
x_{6} \\
x_{9} 
\end{array} 
\right),                          \eqno{(90)}$$
where all blanks stand for zeros.
Therefore,  we can write it as
$$\tilde{A}_{2}
\left(
\begin{array}{c}
\vec{u} \\
\vec{v} \\
\end{array} 
\right)
=\left(
\begin{array}{cc}
0_{3\times 3} & M_{2}^{\dagger}\\
M_{2} & 0_{6\times 6}\\
\end{array} 
\right)
\left(
\begin{array}{c}
\vec{u} \\
\vec{v} \\
\end{array} 
\right)
=\lambda
\left(
\begin{array}{c}
\vec{u} \\
\vec{v} \\
\end{array} 
\right),                                         \eqno{(91)}$$
where $\vec{u}$ and $\vec{v}$ are the $3$ 
and $6$-dimensional vectors, and $0_{3\times 3}$ 
and $0_{6\times 6}$ are zero matrices, 
and
$$M_{2}^{\dagger}=
\left(
\begin{array}{cccccc}
1 & 1 & 1 & 1 & 1 & 1\\
0 & 1 & 0 & 0 & 1 & 0\\
0 & 0 & 1 & 0 & 0 & 1\\
\end{array} 
\right),                                                \eqno{(92)}$$
$$M_{2} =
\left(
\begin{array}{ccc}
1 & 0 & 0  \\
1 & 1 & 0  \\
1 & 0 & 1  \\
1 & 0 & 0  \\
1 & 1 & 0  \\
1 & 0 & 1 
\end{array} 
\right).                                              \eqno{(93)}$$
Here, $\tilde{A}_{2} = PA_{2}P^{-1}$ and 
$\vec{Y}_{2}=(\vec{u}, \vec{v})^{t} = P\vec{X}_{2}$,
respectively.
This is just a interchange of the vector components.
Therefore, if there is no confusion between $\tilde{A}_{2}$ and $A_{2}$,
we can simply write $\tilde{A}_{2}$ as $A_{2}$. 
So, we identify them as the original adjacency matrix $A_{2}$, below.

From Eq.(91), we leads to the following:
$$M_{2}^{\dagger}\vec{v} = \lambda \vec{u},  \
M_{2} \vec{u} = \lambda \vec{v}.                     \eqno{(94)}$$
From this we find by simple algebra
$$M_{2}^{\dagger}M_{2}\vec{u} = {\lambda}^{2} \vec{u},              \
M_{2}M_{2}^{\dagger}\vec{v} = {\lambda}^{2} \vec{v},            \eqno{(95)}$$
where $M_{2}^{\dagger}M_{2}$ is the $3$-dimensional matrix and 
$M_{2}M_{2}^{\dagger}$ the $6$-dimensional matrix.
These are given by
$$M_{2}^{\dagger}M_{2}=
\left(
\begin{array}{cccccc}
1 & 1 & 1 & 1 & 1 & 1\\
0 & 1 & 0 & 0 & 1 & 0\\
0 & 0 & 1 & 0 & 0 & 1\\
\end{array} 
\right)
\left(
\begin{array}{ccc}
1 & 0 & 0  \\
1 & 1 & 0  \\
1 & 0 & 1  \\
1 & 0 & 0  \\
1 & 1 & 0  \\
1 & 0 & 1 
\end{array} 
\right)$$
$$=\left(
\begin{array}{ccc}
6 & 2 & 2 \\
2 & 2 & 0 \\
2 & 0 & 2 \\
\end{array} 
\right),                     \eqno{(96)}$$
$$M_{2}^{\dagger}M_{2}=
\left(
\begin{array}{ccc}
1 & 0 & 0  \\
1 & 1 & 0  \\
1 & 0 & 1  \\
1 & 0 & 0  \\
1 & 1 & 0  \\
1 & 0 & 1 
\end{array} 
\right)
\left(
\begin{array}{cccccc}
1 & 1 & 1 & 1 & 1 & 1\\
0 & 1 & 0 & 0 & 1 & 0\\
0 & 0 & 1 & 0 & 0 & 1\\
\end{array} 
\right)$$
$$=\left(
\begin{array}{cccccc}
1 & 1 & 1 & 1 & 1 & 1\\
1 & 2 & 1 & 1 & 2 & 1\\
1 & 1 & 2 & 1 & 1 & 2\\
1 & 1 & 1 & 1 & 1 & 1\\
1 & 2 & 1 & 1 & 2 & 1\\
1 & 1 & 2 & 1 & 1 & 2\\
\end{array} 
\right).                                          \eqno{(97)}$$

Let us solve the eigenvalue problem of Eq.(96).
In this case, the first eigenvalue problem, 
$M_{2}^{\dagger}M_{2} \vec{u} = \lambda^{2}\vec{u}$,
gives the determinant
$$\tilde{D}_{3}(y)=
\left|
\begin{array}{ccc}
y-6 & -2 & -2 \\
-2 & y-2 & 0 \\
-2 & 0 & y-2 \\
\end{array} 
\right| = 0,                                  \eqno{(98)}$$
where $y = \lambda^{2}$.
By simple algebra, we show the following:
$$\tilde{D}_{3}(y)=
\left|
\begin{array}{ccc}
y-6 & -2 & -2 \\
-2 & y-2 & 0 \\
-2 & 0 & y-2 \\
\end{array} 
\right| $$
$$= 
\left|
\begin{array}{ccc}
y-4 & -2 & -2 \\
0 & y-2 & 0 \\
-2 & 0 & y-2 \\
\end{array} 
\right|
=(y-2)\left|
\begin{array}{ccc}
y-4 & -2 \\
-2 & y-2 \\
\end{array} 
\right|$$
$$=(y-2)\left\{(y-2)(y-4)-4 \right\}
= (y-2)(y^{2} -8y+4).                             \eqno{(99)}$$
Hence, the determinant $\tilde{D}_{3}(y)$ gives
the three eigenvalues of $y=2$, $\sqrt{3}\pm 1$, as expected.

Similarly, the second eigenvalue problem, 
$M_{2} M_{2}^{\dagger} \vec{v} = \lambda^{2}\vec{v}$,
gives the determinant
$$\tilde{D}_{6}(y)=
\left|
\begin{array}{cccccc}
y-1 & -1 & -1 & -1 & -1 & -1\\
-1 & y-2 & -1 & -1 & -2 & -1\\
-1 & -1 & y-2 & -1 & -1 & -2\\
-1 & -1 & -1 & y-1 & -1 & -1\\
-1 & -2 & -1 & -1 & y-2 & -1\\
-1 & -1 & -2 & -1 & -1 & y-2\\
\end{array} 
\right| = 0.                                      \eqno{(100)}$$
Subtracting the first column by the fourth column,
the second column by the fifth column,
and the third column by the sixth column, respectively, 
we obtain
$$\tilde{D}_{6}(y)=
\left|
\begin{array}{cccccc}
y-1 & -1 & -1 & -1 & -1 & -1\\
-1 & y-2 & -1 & -1 & -2 & -1\\
-1 & -1 & y-2 & -1 & -1 & -2\\
-1 & -1 & -1 & y-1 & -1 & -1\\
-1 & -2 & -1 & -1 & y-2 & -1\\
-1 & -1 & -2 & -1 & -1 & y-2\\
\end{array} 
\right| $$
$$=
\left|
\begin{array}{cccccc}
y & 0 & 0 & -1 & -1 & -1\\
0 & y & 0 & -1 & -2 & -1\\
0 & 0 & y & -1 & -1 & -2\\
-y & 0 & 0 & y-1 & -1 & -1\\
0 & -y & 0 & -1 & y-2 & -1\\
0 & 0 & -y & -1 & -1 & y-2\\
\end{array}
\right|.                                                  \eqno{(101)}$$
And as the next step add the first row to the fourth row,
the second row to the fifth row, and
the third row to the sixth row,
we obtain
$$\tilde{D}_{6}(y)=
\left|
\begin{array}{cccccc}
y & 0 & 0 & -1 & -1 & -1\\
0 & y & 0 & -1 & -2 & -1\\
0 & 0 & y & -1 & -1 & -2\\
0 & 0 & 0 & y-2 & -2 & -2\\
0 & 0 & 0 & -2 & y-4 & -2\\
0 & 0 & 0 & -2 & -2 & y-4\\
\end{array}
\right|$$
$$=y^{3}
\left|
\begin{array}{ccc}
y-2 & -2 & -2\\
-2 & y-4 & -2\\
-2 & -2 & y-4\\
\end{array}
\right|
=y^{3}
\left|
\begin{array}{ccc}
y-2 & 0 & -2\\
-2 & y-2 & -2\\
-2 & 2-y & y-4\\
\end{array}
\right|$$
$$=y^{3}
\left|
\begin{array}{ccc}
y-2 & 0 & -2\\
-4 &  0 & y-6\\
-2 & 2-y & y-4\\
\end{array}
\right|
=y^{3}(y-2)
\left|
\begin{array}{cc}
y-2 & -2\\
-4 & y-6\\
\end{array}
\right|$$
$$= y^{3}(y-2)(y^{2}-8y+4).                     \eqno{(102)}$$
Hence, the determinant $\tilde{D}_{6}(y)$ gives
the six eigenvalues of $y=0$ (triple), $2$, $\sqrt{3}\pm 1$, as expected.
We would like to note that the factors seen in Eq.s (99) and (102)
are nothing but the irreducible functions $f_{1}(y)$ and $f_{2}(y)$,
found before.

Let us consider some interesting character in Eq.(91).
Since
$A_{2}\vec{X}_{2} = \lambda\vec{X}_{2}$,
we find that
$$(A_{2})^{2}\vec{X}_{2} = {\lambda}^{2}\vec{X}_{2}, \eqno{(103)}$$
where
$$(A_{2})^{2} =
\left(
\begin{array}{cc}
0_{3\times 3} & M_{2}^{\dagger}\\
M_{2} & 0_{6\times 6}\\
\end{array} 
\right)
\left(
\begin{array}{cc}
0_{3\times 3} & M_{2}^{\dagger}\\
M_{2} & 0_{6\times 6}\\
\end{array} 
\right)$$
$$=
\left(
\begin{array}{cc}
M_{2}^{\dagger}M_{2} & 0\\
0 & M_{2}M_{2}^{\dagger}\\
\end{array} 
\right).                                                  \eqno{(104)} $$
Hence, the square $(A_{2})^{2}$ 
of the original adjacency matrix $A_{2}$ for the DSFN
can be block-diagonal.
Since from Eq.(103) we have
$\{(A_{2})^{2}- {\lambda}^{2}\}\vec{X}_{2} = 0$,
this eigenvalue problem provides the following:
$$\det \left[(A_{2})^{2}- {\lambda}^{2}I_{2}\right]
= \det \left[A_{2} - \lambda I_{2}\right]
\det \left[A_{2} + \lambda I_{2}\right]$$
$$= \left| \det \left[A_{2} - \lambda I_{2}\right]\right|^{2}$$
$$= \det \left[M_{2}^{\dagger}M_{2} - \lambda^{2} I_{2}\right]
\det \left[M_{2} M_{2}^{\dagger} - \lambda^{2} I_{2}\right].       \eqno{(105)}$$
Therefore, the original determinant $D_{2}(\lambda)$ of Eq.(69)
is given by
$$D_{2}(\lambda)
=\left|\det \left[M_{2}^{\dagger}M_{2} - \lambda^{2} I_{2}\right]\right|^{1/2}
\left|\det \left[M_{2} M_{2}^{\dagger} - \lambda^{2} I_{2}\right]\right|^{1/2}$$
$$= (\tilde{D}_{3}(y)\tilde{D}_{6}(y))^{1/2}.                         \eqno{(106)}$$
Substituting Eq.s (99) and (102) into Eq.(106),
we obtain
$$D_{2}(\lambda) = y^{3/2}(y-2)(y^{2}-8y+4)
= \lambda^{3} f_{1}(\lambda)f_{2}(\lambda).                            \eqno{(107)}$$
This reproduces Eq.(76).

Now, we are able to generalize the above procedure to the adjacency matrix $A_{n}$
for the DSFN at any generation. 
In this case the Eq. (91) is generalized to 
$$A_{n}
\left(
\begin{array}{c}
\vec{u} \\
\vec{v} \\
\end{array} 
\right)
=\left(
\begin{array}{cc}
0_{uu} & M_{n}^{\dagger}\\
M_{n} & 0_{vv}\\
\end{array} 
\right)
\left(
\begin{array}{c}
\vec{u} \\
\vec{v} \\
\end{array} 
\right)
=\lambda
\left(
\begin{array}{c}
\vec{u} \\
\vec{v} \\
\end{array} 
\right).                                                     \eqno{(108)}$$
This then yields
$$M_{n}^{\dagger}\vec{v} = \lambda \vec{u},  \ \
M_{n}\vec{u} = \lambda \vec{v}.                               \eqno{(109)}$$
Here
$\vec{u}$ and $\vec{v}$ are the $u$- and $v$-dimensional vectors,
$0_{uu}$ and $0_{vv}$ the $u\times u$ and $v \times v$ zero matrices,
and $M_{n}^{\dagger}$ and $M_{n}$ 
the $u\times v$ and $v \times u$ matrices, respectively,
where $u = 3^{n-1}$ and $v = 3^{n}-3^{n-1}$.
Then, in the same way we have the following:
$$D_{n}(\lambda)
=\left|\det \left[M_{n}^{\dagger}M_{n} - \lambda^{2}I_{n}\right]\right|^{1/2}
\left|\det \left[M_{n} M_{n}^{\dagger} - \lambda^{2}I_{n}\right]\right|^{1/2}$$
$$= \left[\tilde{D}_{3^{n-1}}(y)\tilde{D}_{3^{n}-3^{n-1}}(y)\right]^{1/2},      \eqno{(110)}$$
where $\tilde{D}_{3^{n-1}}(y)$ and $\tilde{D}_{3^{n}-3^{n-1}}(y)$
[$=y^{3^{n-1}}\tilde{D}_{3^{n-1}}(y)$]
are factorized in terms of the irreducible polynomials of Eq.(79).
Therefore, Eq.(107) may reduce to the form of Eq.(75).
Thus, this approach may provide a hint to prove the conjecture 2.
However, the proof is out of scope of this paper
since it is numerically supported as in the section VII.

We would like to mention  that similar approaches
have been applied to the amorphous systems\cite{WT,KE},
topological localization problem\cite{Suther}, 
Fermion number fractionalization\cite{NS}, and
the Hubbard model\cite{Lieb}.

\section{Zero modes and Index theorem in the deterministic scale-free network}
Let us consider {\it zero modes} (i.e., the eigenvectors having $\lambda = 0$) in the spectrum.
The zero modes are given by substituting $\lambda = 0$ into Eq.(107)
such that
$$M_{n}^{\dagger}\vec{v} = 0,  \ \
M_{n}\vec{u} = 0.                                                  \eqno{(111)}$$
The number of zero modes given by $M_{n}\vec{u} = 0$
(or equivalently, the number of zero modes given 
by the matrix $M_{n}^{\dagger}M_{n}$)
is the dimension of the null space of $M_{n}$.
This is sometimes called the dimension of the kernel of $M_{n}$,
that is simply written as $\dim[\ker M_{n}]$.
And the number of zero modes given by $M_{n}^{\dagger}\vec{v} = 0$
(or equivalently, the number of zero modes given by the matrix $M_{n}M_{n}^{\dagger}$)
is the dimension of the null space of $M_{n}^{\dagger}$.
This is sometimes called the dimension of the kernel of $M_{n}^{\dagger}$,  
simply written as $\dim[\ker M_{n}^{\dagger}] = \dim[co \ker M_{n}]$\cite{Nakahara}.
Therefore, in the above example of $n=2$,
$\dim[\ker M_{2}] = 3$
and
$\dim[\ker M_{2}^{\dagger}] = 0$.
This would give a relation:
$$Ind(M_{2}) \equiv \mbox{the number of zero modes}$$ 
$$= \dim[\ker M_{2}] - \dim[\ker M_{2}^{\dagger}]  = 3.             \eqno{(112)}$$
This quantity $Ind(M_{n})$ is called the {\it index} of $M_{n}$.
And the relation that the number of zero modes coincides
with the difference between $\dim[\ker M_{n}]$ and $\dim[\ker M_{n}^{\dagger}]$ 
is called the {\it index theorem}\cite{Nakahara}.

This can be generalized to the adjacency matrix $A_{n}$ 
for the DSFN in the $n$-th generation such that we have
$$Ind(M_{n}) = \dim[\ker M_{n}] - \dim[\ker M_{n}^{\dagger}].             \eqno{(113)}$$
As we have obtained Eq.(78), for this case we obtain
$$\dim[\ker M_{n}] = 3^{n-1}, \ \  \dim[\ker M_{n}^{\dagger}]  = 0.         \eqno{(114)}$$
Hence, we derive the following index theorem for the DSFN:
$$Ind(M_{n}) = \dim[\ker M_{n}] - \dim[\ker M_{n}^{\dagger}]  = 3^{n-1}.         \eqno{(115)}$$
As we have discussed before, this number is just the number of $\lambda = 0$,
localized at the smallest leaf nodes.

The proof of Eq.(114) is quite simple.
For the DSFN, the dimension of the matrix $M_{n}^{\dagger}M_{n}$
is $3^{n-1}$, which is the number of the hub nodes.
And the dimension of the matrix $M_{n}M_{n}^{\dagger}$
is $3^{n}-3^{n-1}$, which is the number of the rim nodes.
Then, we always find that there are no zero modes
in $M_{n}^{\dagger}M_{n}$ in our DSFN.
That is, $ \dim[\ker M_{n}^{\dagger}]  = 0$.
Therefore, the null space of
$M_{n}M_{n}^{\dagger}$ is given by the difference between
the dimension of the matrix $M_{n}M_{n}^{\dagger}$ and
that of the matrix $M_{n}^{\dagger}M_{n}$.
This is $3^{n}-3^{n-1} - 3^{n-1} = 3^{n-1}$.
Hence, $\dim[\ker M_{n}] = 3^{n-1}$.

\section{The nature of the maximum eigenvalue}
Let us consider the nature of the maximum eigenvalue.
As we studied before, the matrix $M_{n}^{\dagger}M_{n}$
absolutely determines the spectrum of $\lambda > 0$,
while the matrix $M_{n}M_{n}^{\dagger}$ determines
the zero modes of $\lambda = 0$ as well as the spectrum.
Therefore, in order to determine the spectrum
we need study the matrix $M_{n}^{\dagger}M_{n}$.

Let us see some examples for this matrix.
The matrix for $n = 2$ is given in Eq.(96).
We also give the matrix for $n = 3$ as
$$M_{3}^{\dagger}M_{3}=
\left(
\begin{array}{ccccccccc}
14 & 2 & 2 & 4 & 2 & 2 & 4 & 2 &2\\
2 & 2 & 0 & 0 & 0 & 0 & 0 & 0 & 0\\
2 & 0 & 2 & 0 & 0 & 0 & 0 & 0 & 0\\
4 & 0 & 0 & 6 & 2 & 2 & 0 & 0 & 0\\
2 & 0 & 0 & 2 & 2 & 0 & 0 & 0 & 0\\
2 & 0 & 0 & 2 & 0 & 2 & 0 & 0 & 0\\
4 & 0 & 0 & 0 & 0 & 0 & 6 & 2 & 2\\
2 & 0 & 0 & 0 & 0 & 0 & 2 & 2 & 0\\
2 & 0 & 0 & 0 & 0 & 0 & 2 & 0 & 2\\
\end{array} 
\right).                                             \eqno{(116)}$$
Here, we find that the diagonal elements 
of the matrix $M_{3}^{\dagger}M_{3}$
are just the maximum degrees of nodes 
(i.e., the numbers of the most connected links)
in the networks up to the third generation.
The maximum diagonal element is the maximum degree of nodes,
which is the degree of the root.
Hence, it is $14$ for $n = 3$ 
[Obviously, it is $6$ for $n = 2$ as in Eq.(96)].

We find that this is always true for the network of the $n$-th generation.
We find that the diagonal elements of the matrix $M_{n}^{\dagger}M_{n}$
are just the maximum degrees of nodes for the networks up to the $n$-th generation.
Hence the maximum diagonal element is the maximum degrees of node,
which is the degree of the root.
It is $k_{max} = 2(2^{n} -1)$
from Eq.(7).

Let us first derive the lower bound of the maximum eigenvalue, $\lambda_{max}$.
In mathematics, we have the following theorem\cite{Ortega}:
\newtheorem{theorem}{Theorem}
\begin{theorem}
Denote by $H$ a non-negative definite matrix.
Denote by $\vec{v}$ a positive definite symmetric matrix.
Define a matrix $A$ by $A = H + V$.
Suppose that the eigenequations 
$H|\psi_{i} \rangle = h_{i} |\psi_{i}\rangle$ 
and
$A|\psi_{i} \rangle = a_{i} |\psi_{i}\rangle$
($i = 1, 2, \cdots, l$)
are known such that the eigenvalues satisfy
$h_{1} \ge h_{2} \ge \cdots \ge h_{l}$
and
$a_{1} \ge a_{2} \ge \cdots \ge a_{l}$.
And assume that
$\langle\psi_{i}| V |\psi_{i}\rangle \ge 0$.
Then, the following holds true
$$h_{s} \le  a_{s} \ \ \mbox{for $s = 1, 2, \cdots, l$}.      \eqno{(117)}$$
\end{theorem}
We omit the proof here, since it is given in the literature\cite{Ortega}.

Let us apply this theorem to the matrix $M_{n}^{\dagger}M_{n}$.
We now denote the matrix $M_{n}^{\dagger}M_{n}$  by $A$ in the theorem.
Denote by $H$ the matrix having the diagonal matrix components of $M_{n}^{\dagger}M_{n}$ 
and denote by $\vec{v}$ the matrix having the off-diagonal matrix components of $M_{n}^{\dagger}M_{n}$.
Therefore, $M_{n}^{\dagger}M_{n}$ satisfies the conditions for $A$ in the theorem.
The eigenvalue of the matrix $A$ is now $a_{i}=\lambda_{i}^{2}$
and $l = 3^{n-1}$.
Therefore, we can prove the following theorem for the DSFN.
\begin{theorem}
$$h_{s} \le \lambda_{s}^{2}  \ \ \mbox{for $s = 1, 2, \cdots, 3^{n-1}$}.  \eqno{(118)}$$
\end{theorem}
Since the maximum eigenvalue of $H$ is now $k_{max}$, 
we have
$h_{1} =  k_{max}$.
Thus, we finally end up with the following theorem for the DSFN.
\begin{theorem}
$$\sqrt{k_{max}} = \sqrt{2^{n+1} -2} \le \lambda_{max}.  \eqno{(119)}$$
\end{theorem}

We have also investigated this theorem numerically in Fig.4.
It shows that the theorem is valid for the DSFNs of the generations up to $n=8$.
Thus, this supports the theorem.

\begin{figure}[h]
\includegraphics[scale=0.6]{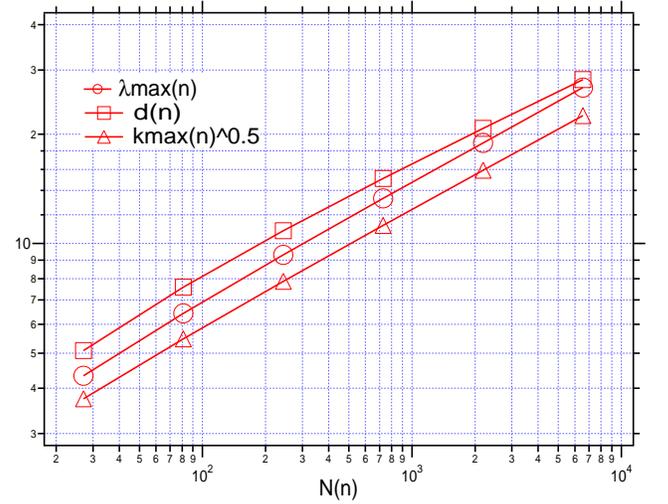}
\caption{
(Color online)
The growth of the maximum eigenvalue, $\lambda_{max}$ 
and its lower and upper bounds in the DSFN.
Here $\lambda_{max}$ (circle), the lower (triangle), and 
upper (square) bounds are shown for the DSFNs of the $n$-th generations 
up to $n=8$, respectively.
} 
\end{figure}

Let us next derive the upper bound of the maximum eigenvalue, $\lambda_{max}$.
To do so, let us first consider a particular property of the matrix, $M_{n}$.
As is seen from the matrix such as Eq.(93),
the $i$-th column vector of $M_{n}$ are a vector whose components are $0$ or $1$
such that the total number of $1$ in the column counts the order $k_{i}$ of the $i$-th hub node.
Denote this vector by $\vec{k}_{i}$.
Then we find 
$$\vec{k}_{i}\cdot \vec{k}_{i} = k_{i},                          \eqno{(120)}$$
where by definition $k_{1} >  k_{2} \ge \cdots \ge  k_{3^{n}}$,
since $k_{1}$ is the order of the root (i.e., the most connected hub).
Then, the matrix can be represented as
$$M_{n} = (\vec{k}_{1}, \vec{k}_{4}, \cdots, \vec{k}_{3^{n}-2}), \eqno{(121)}$$
where the suffices run over all the hub (i.e., root) nodes.
Now, we can represent $M_{n}^{\dagger}M_{n}$ in general in the following:
$$M_{n}^{\dagger}M_{n} =
\left(
\begin{array}{cccc}
\vec{k}_{1}\cdot \vec{k}_{1} & \vec{k}_{1}\cdot \vec{k}_{4} & \cdots &\vec{k}_{1}\cdot \vec{k}_{3^{n}-2}\\
\vec{k}_{1}\cdot \vec{k}_{4} & \vec{k}_{4}\cdot \vec{k}_{4} & \cdots & \vec{k}_{4}\cdot \vec{k}_{3^{n}-2} \\
\vdots & \vdots & \ddots & \vdots\\
\vec{k}_{1}\cdot \vec{k}_{3^{n}-2} & \vec{k}_{2}\cdot \vec{k}_{3^{n}-2} & \cdots & \vec{k}_{3^{n}-2}\cdot \vec{k}_{3^{n}-2}\\
\end{array} 
\right),                                                       \eqno{(122)}$$
which is a $3^{n-1}\times 3^{n-1}$ symmetric matrix.
Therefore, the trace of this matrix is the total number of links
such that
$$tr \left(M_{n}^{\dagger}M_{n} \right)  
=\sum_{i \in root} \vec{k}_{i}\cdot \vec{k}_{i}
=\sum_{i \in root} k_{i}$$
$$=\sum_{k \in  root} k P(k) = L(n).                      \eqno{(123)}$$

We now need the following theorem known as 
the Perron-Frobenius theorem in linear algebra\cite{CLV,Ortega}.
\begin{theorem}
Suppose that an $n\times n$ symmetric matrix $A$
has all non-negative entries $a_{ij} \ge 0$.
Then this satisfies an eigenequation
$A|\psi_{i} \rangle = a_{i} |\psi_{i}\rangle$.
For any positive constants $c_{1}, c_{2}, \dots, c_{n}$,
the maximum eigenvalue $a_{max}(A)$ satisfies
$$a_{max}(A) \le \max_{1 \le i \le n} 
\left\{ \sum_{j=1}^{n} \frac{c_{j} a_{ij}}{c_{i}}\right\}.  \eqno{(124)}$$
\end{theorem}
We omit the proof here since this is very well-known\cite{CLV,Ortega}.

Let us apply this theorem to our problem.
We now regard $A$ as $M_{n}^{\dagger}M_{n}$
such that $a_{ij} = \vec{k}_{i}\cdot \vec{k}_{j}$
and
$a_{i} = \lambda_{i}^{2}$.
Then, from the theorem we find the following:
$$\lambda_{max}^{2}(M_{n}^{\dagger}M_{n}) \le 
\max_{i \in root} 
\left\{ \sum_{j \in root} \frac{c_{j} 
\vec{k}_{i}\cdot \vec{k}_{j}}{c_{i}}\right\}.            \eqno{(125)}$$
From this we also find
$$\lambda_{max}^{2}(M_{n}^{\dagger}M_{n}) \le 
\max_{i \in root} 
\left\{ \sum_{j\in root} \frac{c_{j} 
\vec{k}_{i}\cdot \vec{k}_{j}}{c_{i}}\right\}$$
$$\le \max_{i \in root} 
\left\{ \sum_{j\in root} \left|\frac{c_{j} 
\vec{k}_{i}\cdot \vec{k}_{j}}{c_{i}}\right| \right\}$$
$$\le \max_{1 \le i \le n} 
\left\{ \sum_{j\in root} \vec{k}_{i}\cdot 
\vec{k}_{j}\left|\frac{c_{j}}{c_{i}}\right| \right\}.     \eqno{(126)}$$
Since $|c_{j}/c_{i}| < 1$, then we have
$$\lambda_{max}^{2}(M_{n}^{\dagger}M_{n}) \le  
\max_{i \in root} \left\{ \sum_{j\in root} 
\vec{k}_{i}\cdot \vec{k}_{j} \right\}.                   \eqno{(127)}$$
Hence, we have the following theorem for the DSFN.
\begin{theorem}
$$\lambda_{max}^{2}(M_{n}^{\dagger}M_{n}) 
\le \sum_{j\in root} \vec{k}_{1}\cdot \vec{k}_{j} \equiv b.   \eqno{(128)}$$
\end{theorem}
Since
$\sum_{j\in root} \vec{k}_{1}\cdot \vec{k}_{j} \le 
\sum_{j\in root} \vec{k}_{j}\cdot \vec{k}_{j} = L(n)$
together with Eq.(6),
we end up with the following theorem for the DSFN.
\begin{theorem}
$$\lambda_{max}(M_{n}^{\dagger}M_{n}) \le 
\sqrt{L(n)} = \sqrt{2(3^{n} - 2^{n})}.                       \eqno{(129)}$$
\end{theorem}
From theorem 3 and theorem 6, we finally derive 
the following theorem for the DSFN.
\begin{theorem}
The maximum eigenvalue of the adjacency matrix 
of the DSFN is bounded as
$$a \le \lambda_{max} \le b \le c,                          \eqno{(130)}$$
where $a = \sqrt{2(2^{n} -1)}$, $c=\sqrt{2(3^{n} - 2^{n})}$, 
and
$b = \sum_{j\in root} \vec{k}_{1}\cdot \vec{k}_{j}$.
\end{theorem}
Although there is no name on the quantity $b$,
this is a much better bound than $c$, 
the square root of the total number of links.
We note that the above theorem is consistent with
Conjecture 3 that asserts $\lambda_{max} \rightarrow 2^{n/2}$
as $n \rightarrow \infty$.

We finally make a comment on the Chung, Lu and Vu's theorem\cite{CLV}.
As introduced in Introduction, since the DSFN has 
the exponent of $\gamma = \ln3/\ln2 < 2.5$,
this system may belong to the (C2) case.
Therefore, 
if we apply their theorem to our system of the DSFN,
then it leads us to the following:
$$\sqrt{k_{max}} \le \lambda_{max} \le \tilde{d},         \eqno{(131)}$$
where $\tilde{d} \propto (4/3)^{n} =(1.33\cdots)^{n}$ from Eq.(49).
On the other hand, our theorem provides us  
the lower bound
$a \propto 2^{n/2}=(1.414\cdots)^{n}$
and 
the upper bound $c \propto 3^{n/2}=(1.732\cdots)^{n}$.
From this, we find that the Chung, Lu and Vu's theorem\cite{CLV}
holds true until about $n = 12$.
However, beyond $n = 12$ the role between
$\tilde{d}$ and $\sqrt{k_{max}}$ is switched.
So, their theorem can be violated above $n=12$,
although strictly speaking, their theorem should be
applied to the case of $ 2 < \gamma < 2.5$.
We show this behavior using the above exact expressions in Fig.5.
As is shown in Fig.4,
the numerical value of $\lambda_{max}$ is sandwiched between the upper bound of $\tilde{d}$ 
and the lower bound of $\sqrt{k_{max}}$.
Therefore, it follows the Chung, Lu, and Vu's theorem within our calculation
up to $n = 8$.
Unfortunately, the numerical value of $\lambda_{max}$
is not available above $n = 8$ because of the ability of our computer power.
So, we cannot say anything about where it is located above $n = 8$, so far.
We may expect that it is still sandwiched between them
although the role is switched.

In this way, our rigorous approach provides a concrete example
for investigating the validity of the Chung, Lu and Vu's theorem\cite{CLV}.
This is an advantage of our theory.

\begin{figure}[h]
\includegraphics[scale=0.6]{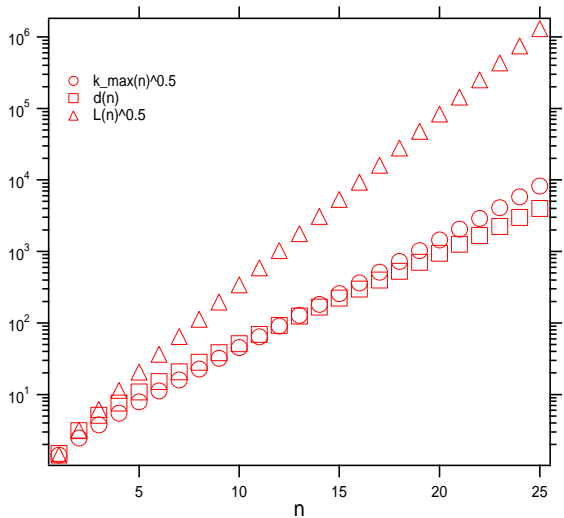}
\caption{
(Color online)
The growth of the square root of the total number of links $\sqrt{L(n)}$, 
the maximum order of nodes $\sqrt{k_{max}}$,
and
the maximum degree $\tilde{d}(n)$. 
Clearly, we see that
$\sqrt{k_{max}}$ is smaller than $\tilde{d}(n)$
but very close to  it under $n=12$.
However,  
$\sqrt{k_{max}}$ exceeds  $\tilde{d}(n)$
about $n=12$.
} 
\end{figure}

\section{conclusions}
In conclusion, we have intensively studied the DSFN that was first studied by
Barab\'{a}si, Ravasz, and Vicsek\cite{BRV}.
We have first studied the geometry of the network
and presented the exact numbers of nodes and degrees.
From these we were able to calculate the exact scaling exponents
for the hub and rim nodes, where we have shown that
the scaling nature of the hubs is a scale-free law but
that of the rims is not so but the exponential.
This is obtained only when we have known the exact numbers of the nodes and degrees.
So, the scaling behaviors of the hub and rim nodes are different.
In some sense we may say that the DSFN is more like a multifractal.
This nature has not found yet by Barab\'{a}si, Ravasz, and Vicsek\cite{BRV}.

Second, we have analytically calculated the exact number of 
the second order average degree $\tilde{d}$ [see Eq.(49)], 
using the numbers of the nodes and degrees.
This quantity has been frequently used in the criterion for 
investigating  the bounds of the maximum eigenvalue $\lambda_{max}$
of a SFN as Chung, Lu, and Vu\cite{CLV} have been emphasizing.
However, there has been no example of a rigorous result.
So, this would be a first example of the analytically calculated 
quantity, $\tilde{d}$.

Third, we have numerically calculated the spectra of the adjacency
matrix $A_{n}$ for the network up to the $n = 7$-th generation.
From this we have counted the exact numbers of degeneracies in the spectrum,
and have shown that the density of states looks like a fractal, where
there is a large peak at the center of the spectrum of $\lambda = 0$
[see Fig.2 and Fig.3].
Such degeneracies are related to the degree
of the localization of the state (i.e., the eigenvector).
As the eigenvalue becomes larger the state becomes broader (i.e., delocalized), 
where the minimum eigenvalues of $\lambda = 0$ are localized 
on the minimum connected rim nodes.
The envelope of the density of states in the DSFN is somehow similar to that
of the spectral shape in the AB-model\cite{Barabasi}.
So, we expect that the origin of localization in the AB-model may be
the same or similar to that in the DSFN that we have studied here. 

Fourth, we have discussed the nature of the adjacency matrix for the network.
We have shown that there is a recursive structure of the
determinant, which induces to the complex structure of degeneracy.
The sequence of the irreducible polynomial seems very important.
This would be related to a new type of functions.
The roots of the polynomials and their irreducibility may be 
a problem of the Sturm theorem in algebra.
Therefore, this will be an interesting problem for mathematicians.

Fifth, we have shown that there is a hidden symmetry in the adjacency matrix.
This is the consequence of a bipartite structure of the network.
From this nature, the adjacency matrix is decomposed into the off-diagonal type.
Then we have proved an index theorem for the network for the first time.
This theorem has enabled us to count the number of zero modes exactly.
This type of theory is very well-known in the quantum field theory\cite{NS,Nakahara}.
It is known as the supersymmetric structure.
And it has been also known in solid state physics from a long time ago\cite{WT,KE,Suther,NS,Lieb}.
The number of zero modes is related to the number of localized states in the system.
Therefore, we expect that this may be the case in other network systems.
This will be a very important problem.

Finally, we have investigated the maximum eigenvalue, $\lambda_{max}$ in the DSFN.
We have shown that $\lambda_{max}$ is bounded by the lower and upper bounds
that are given analytically [see Theorem 7].
These bounds grow as fast as exponential as the network grows.
Hence, by the theorem, $\lambda_{max}$ grows exponential as well.
This might be also a first example for the determination of the exact bounds.
In the standard networks such as the random networks\cite{Barabasi},
the maximum eigenvalue cannot grow so fast as the network grows.
And also, as in solid state physics, networks in most of physical systems 
provide the so-called energy band that is a spectrum with a finite region.
This is due to the topology of the finite coordination number 
of atoms in the network of the lattice structure.
So, in order to elucidate the difference between the SFNs and other networks
the growth of the maximum eigenvalue is an important signature.
As was studied by many authors\cite{FFF,Barabasi,AB,FDBV,GKK,DGMS},
the behavior of $\lambda_{max}$ in the AB-model
is believed to be proportional to $\sqrt{k_{max}}$.
Therefore, since $k_{max}$ is propotional to $\sqrt{N}$,
hence, $\lambda_{max} \propto N^{1/4}$.
To see whether or not this is true in an arbitrary SFN
and to know how general it is, Chung, Lu, and Vu\cite{CLV}
established the theorem.
However, as we have shown, $\lambda_{max}$ in the  DSFN 
that we have studied in this paper seems to escape from the
region of their theorem. 
Therefore, our example will be a counter example for their theorem.
Whether or not this is true will be an interesting problem for further study.

Thus, we believe that our theory presented in this paper
gives the first rigorous example in the SFN theory,
where most of all quantities in the network theory are analytically obtained.
In this context,  we would like to call the DSFN
the exactly solvable SFN model.

\acknowledgments
One of us (K. I.) would like to thank Kazuko Iguchi
for her financial support and encouragement.

\end{document}